\documentclass[a4paper,11pt]{article}
\usepackage{amssymb} 
\usepackage{epsfig}
\usepackage{natbib}
\bibpunct{(}{)}{;}{a}{,}{;}

\usepackage{latexsym}

\newcommand{\Rm}{R_m}

\newcommand{\upartial}{\partial}

\newcommand{\bnabla}{\mbox{\boldmath $\nabla$}}
\renewcommand{\vec}[1]{{\bf #1}}

\font\sls = cmssi10 
\newcommand{\mat}[1]{\mbox{\sls #1}}

\newcommand{\rf}[1]{(\ref{#1})}
 
\newcommand{\laplace}{\nabla^2}  
\newcommand{\cross}{\wedge}

\newcommand{\curl}{\bnabla \wedge}

\newcommand{\pd}[1]{\upartial_{#1}}

\newcommand{\ex}{{\mathrm e}}

\title{
   Kinematic dynamo action in a sphere:
   Effects of periodic time-dependent flows on solutions with axial dipole 
   symmetry.
}

\author{
   Ashley P. Willis\ \ and\ \ David Gubbins
\\ \it
   School of Earth Sciences, University of Leeds, LS2 9JT, UK
}

\date{\today}

\begin{document}
   
\maketitle

{\abstract\noindent
Choosing a simple class of flows, with characteristics that may be
present in the Earth's core, we study the ability to generate
a magnetic field when the flow is permitted to oscillate periodically
in time.  The flow characteristics are parameterised by 
$D$, representing a differential rotation, 
$M$, a meridional circulation, and
$C$, a roll component characterising convective rolls.
The dynamo action of all solutions with fixed parameters
(steady flows)
is known from previous studies.
Dynamo action is sensitive to these flow parameters
and fails spectacularly for much of the
parameter space where magnetic flux is concentrated into small 
regions, leading to high diffusion. 
In addition, steady flows generate only steady
or regularly reversing oscillatory fields
and cannot therefore reproduce 
irregular geomagnetic-type reversal behaviour.
Oscillations of the flow are introduced by varying the flow
parameters in time, defining a closed orbit in the space $(D,M)$.
When the frequency of the oscillation is small, 
the net growth rate of the magnetic field over one period
approaches the average of the growth rates for steady flows
along the orbit.
At increased frequency time-dependence appears to smooth out flux
concentrations, often enhancing dynamo action.
Dynamo action can be impaired, however, when flux 
concentrations of opposite signs occur close together as
smoothing destroys the flux by cancellation. 
It is possible to produce geomagnetic-type reversals by making the 
orbit stray into a region where the steady flows generate 
oscillatory fields.  In this case, however,
 dynamo action was not found to be
enhanced by the time-dependence.
A novel approach is taken to solving the time-dependent eigenvalue 
problem, where by combining Floquet theory with
a matrix-free Krylov-subspace method we 
avoid large memory requirements for storing the matrix required by 
the standard approach.
\\ \\
{\it Keywords:} Kinematic dynamos; Time dependent stability; 
 Geomagnetism; Floquet theory; Eigenvalue problems.
}

\section{Introduction}
Unlike the Sun or Jupiter, the Earth's dynamo 
runs on a tight heat budget and may therefore rely on a 
significant large-scale component to the flow.  
In addition to the dominant dipole component of the geomagnetic
field, there is some observational evidence to suggest the
field contains a persistent non-axisymmetric component
\citep{gubbins93}.  
Both may be indicative of
a steady component to the underlying core flow. 
This prompted a kinematic study of dynamo action from a class 
of large-scale steady candidate core-flows 
(\citet{gubbins00a}, \citet{gubbins00b}, \citet{gubbins02}
hereafter referred to as papers I--III). 
Steady flows, particularly if they contain stagnation points, 
tend to concentrate magnetic flux into small regions with large 
energy loss due to diffusion. 
Often, increasing the flow speed to overcome diffusive loss simply 
results in more concentrated flux and faster decay. 
On the other hand, chaotic flows appear to make better dynamos, 
perhaps because the mixing properties of the flow prevents
permanent flux concentration and exponential separation of 
neighbouring particles in the flow lead to stretching of the 
magnetic field \citep{brummell98}.
The dynamo mechanisms of complicated chaotic flows
are difficult to understand.  We are therefore motivated to first
study the effects of simple flows that fluctuate about a steady mean.
Although the Earth's dipole has persisted for a long time, 
secular variation including excursions of the magnetic pole may
indicate that fluctuations of the large-scale flow are 
present in the Earth's core.

Kinematic theory ignores the nonlinearity of back-reaction by the 
magnetic field on the flow,
and considers only the time evolution of the magnetic field 
$\vec{b}$ as governed by the induction equation
\begin{equation}
   \label{eq:indeqn}
   \partial_t \vec{b} = \Rm \curl ( \vec{u} \cross \vec{b} ) 
   + \laplace \vec{b} .  
\end{equation}
The induction equation has been non-dimensionalised with the
timescale for magnetic diffusion, the length scale
$d$, the radius of the sphere,
and in this work the 
velocity $\vec{u}$ is normalised such that the
magnetic Reynolds number $R_m$ is unity for a flow
of unit kinetic energy.  For a given steady flow the induction 
equation is linear in $\vec{b}$ and has eigenfunction
solutions of the form
$ 
   \vec{b}(r,\theta,\phi;t) = \ex^{\sigma t}\,\vec{B}(r,\theta,\phi) .
$ 
Dynamo action is established if $\Re({\sigma})>0$. 
This simple test is the major
advantage of the kinematic approach.  The
alternative is to integrate in time the nonlinear problem
for both the velocity and magnetic field
until one is convinced the magnetic field will not ultimately decay;
this is expensive and the results can be uncertain. 
The advantage remains, however, when the flow varies in time
but is periodic;
Floquet theory gives an eigenvalue problem for the growth rate. 

\citet{backus58} was first to show kinematic dynamo action by 
a time-dependent flow.
His dynamo employed periods of stasis while high harmonics in the field
decayed, enabling him to establish convergence of the solution. 
Time dependence
may even lead to dynamo action when no single snapshot of the flow can
generate magnetic field on its own. 
Magnetic fields can grow during an initial transient period
under the influence of a 
subcritical steady flow, the most familiar example being 
the production of toroidal field from the action of differential rotation
on poloidal field. 
If the induction equation were self-adjoint and its eigenfunctions 
orthogonal it would be a simple matter to prove that all such transients 
decay; it is the non-normal property that allows transients to grow.
The initial fields which optimise transient growth for flows in a sphere,
including one of the flows here, have been studied by \citet{livermore04}.  
Unfortunately, if the flow is steady the field eventually dies away 
and only the slowest decaying mode remains.
If the flow is permitted to be time dependent, however,
once a transient field associated with the initial flow has grown, 
a change in the flow can encourage further growth.
In plane-layer flow it has been shown that by repeatedly switching 
the orientation of the flow it is possible to take
advantage of these transients \citep{gog99}, 
and to find dynamo action where each flow in isolation does not 
dynamo kinematically.

Another reason to extend the studies to time-dependent flows
is that steady flows cannot 
account for the irregularity of geomagnetic reversals.
The induction equation is linear with eigenfunctions $\vec{b}$ that
change with $t$ only in magnitude, when $\Im(\sigma)=0$, 
or oscillatory solutions 
that reverse polarity with fixed period $2\pi/\Im(\sigma)$.
Geomagnetic-type reversals require changes in the flow. 
\citet{sarson99} described 
irregular reversals that occurred in simulations with their
$2\frac1{2}$-dimensional model.  
The mechanism could be interpreted kinematically, and reversals were
observed to occur when fluctuations in the flow lead to a 
reduced meridional circulation.
More recently \citet{wicht04} studied reversals in a fully self-consistent
but quasi-periodic system.
The reversal mechanism they proposed also appeals largely to kinematic 
principles and appears to reverse with approximately fixed
period even when nonlinearity through the Lorentz force is 
omitted.


The class of steady flows explored in I--III was originally
prescribed by \citet{kumar75} and, with parameters chosen to mimic 
flows near the limit of \citet{braginsky64}, was shown to be capable
of dynamo action.  Dependence of the dynamo on a much wider 
range of parameter values was later found in I.
The Kumar--Roberts flow is confined to the sphere of unit radius, 
the exterior of which is assumed to be perfectly insulating.
Three components of the flow represent
a differential rotation, a meridional circulation and a 
convective overturn,
\begin{equation}
   \vec{u} = 
   \epsilon_0 \vec{t}_1^0 + 
   \epsilon_1 \vec{s}_2^0 + 
   ( \epsilon_2 \vec{s}_2^{2c} + 
     \epsilon_3 \vec{s}_2^{2s} ) .
\end{equation}
Following the nomenclature detailed in I,
the $\epsilon_i$ are constrained such that $\epsilon_2=\epsilon_3$ and
the kinetic energy of the flow is unity.
The flow is parameterised by $(D,M)$-space, where 
$D=D(\epsilon_0)$, $M=M(\epsilon_1)$ and $|D|+|M|\le 1$.
The parameters $D$ and $M$ are measures of the
differential rotation and meridional circulation respectively.

For a steady forcing flow,  writing 
$\vec{b}(t)=\ex^{\sigma t}\vec{B}$, where $\vec{B}$ is independent of $t$,
\rf{eq:indeqn}
can be expressed as the eigenvalue problem
\begin{equation}
   \label{eq:eigprobfnsteady}
   \sigma \, \vec{B} = {\mathcal F} \, \vec{B} .
\end{equation}
In paper I dynamo action was established for 
approximately half the $(D,M)$-space
(Fig.~\ref{fig:diamond}).
\citet{sarson96} and III
found a number of oscillatory solutions for steady flows
in a region which corresponds to the dynamo wave solutions of the 
$\alpha\omega$ equations in the Braginsky limit ---
$|D|\to 1$ in a manner such that $1-|D|^2=c\,|M|$ where $c$ is a constant.
The oscillatory region was found in I
to extend only for a very narrow range in $M$, shown schematically 
in Fig.~\ref{fig:diamond}. The majority of solutions are steady.
Given the narrow range for $M$, it is apparent that 
only a small degree of meridional circulation is required to stabilise 
the field to steady solutions. On the other hand, the existence of 
oscillatory solutions for low $M$ appears to be a fairly robust feature
as the range in $D$ for which they exist is large, and extends well beyond 
the limit of Braginsky.  

In this work, the exploration above is extended to the
dynamo action of flows with $D=D(t)$ and $M=M(t)$ periodic in time, 
with a given period $T$. 
The induction equation \rf{eq:indeqn} can be written as
\begin{equation}
   \label{eq:indop}
   \pd{t} \vec{b} = {\mathcal F}(t) \, \vec{b},
\end{equation}
with periodic forcing
   ${\mathcal F}(T+t) = {\mathcal F}(t)$.
It follows from Floquet's theorem (see \S\ref{sect:nummeth})
that solutions may be written in the form
$\vec{b}(T+t)=\ex^{\sigma_1 T}\,\vec{b}(t)$
where the real part of $\sigma_1$ is the net growth rate over one
period.  Setting $\vec{b}(t)=\ex^{\sigma_1 t}\,\vec{B}(t)$, so that
$\vec{B}(T+t)=\vec{B}(t)$,
substitution into \rf{eq:indop} defines the 
eigenvalue problem for $\vec{B}$,
\begin{equation}
   \label{eq:eignonst}
   \sigma_1 \, \vec{B} = \left( {\mathcal F} - \pd{t} \right) \vec{B} .
\end{equation}
The critical magnetic Reynolds number for which the field is marginally
stable, $\Re(\sigma)=0$, is denoted $\Rm^c$.

Both the steady and non-steady 
eigenvalue problems \rf{eq:eigprobfnsteady} and \rf{eq:eignonst}
permit solutions for four linearly independent spatial
symmetries, axial dipole, axial quadrupole, equatorial dipole
and equatorial quadrupole.  Symmetry selection in the steady case
was studied in II. Here only the geophysically interesting axial dipole
symmetry will be considered.
\section{Numerical method}
\label{sect:nummeth}

Steady flows have been studied using extensions of the method first
developed by \citet{bullard54}. Toroidal and poloidal potentials for
the magnetic field are expanded in
spherical harmonics, with 
truncation at degree $L$.  A finite difference scheme is applied 
on $N_r$ points in the radial dimension 
leading to 
the discretised eigenvalue problem
\begin{equation}
   \label{eq:eigprobmat}
   \sigma \, \vec{B} = \mat{E} \, \vec{B} .
\end{equation}
The matrix $\mat{E}$ has dimensions $N_rN_h\times N_rN_h$, where after
symmetry considerations the number of harmonics $N_h \sim \frac1{2} L^2$.
As the finite difference scheme only connects neighbouring points, 
$\mat{E}$ is block banded where each block has size $N_h\times N_h$.
Eigenvectors are then 
calculated by either by inverse iteration or by the Implicitly
Restarted Arnoldi Method (IRAM) on the inverse.  
Due to the performance of both methods with respect to the 
distribution of the eigenvalues, both operate on the
inverse and require the (banded) LU factorisation of $\mat{E}$.
Memory requirements scale like several times $N_h^2 N_r$, depending on
the stencil size of the finite difference approximation.  
Solutions have generally been calculated with second order differences,
and $L$ not much larger than twenty. The storage requirement for the large
matrix is the limiting factor for the calculation.

For the time-dependent eigenvalue problem \rf{eq:eignonst} with
the same spatial representation, applying 
a Fourier expansion in time introduces at least another factor
$N_t$ to the storage requirements.  This can be minimised by
permitting only sinusoidal forcings, but due to the structure of the 
matrix memory requirements are prohibitive
with respect to calculation of the 
LU factorisation (a few times $N_h^2 N_r^2 N_t$).
Storage is a significant difficulty in multiplying by the inverse
or in calculating the inverse of a suitable preconditioner for the
time-dependent problem.


Instead we have adopted a method that does not require storage of the
matrix, which we call the matrix-free Krylov subspace method.
It is an adaptation of a 
method used to find steady solutions of the Navier-Stokes equations
by \citet{edwards94}.  Periodicity of the flow is incorporated in
the following manner \citep{verhulst96}.
Writing the discrete form of (\ref{eq:indop}) as
\begin{equation}
   \label{eq:indopdesc}
   \pd{t}\vec{b} = \mat{F}(t) \vec{b},
\end{equation}
the matrix $\mat{G}(t)$ satisfying
$ 
   \pd{t}\mat{G}(t) = \mat{F}(t)\,\mat{G}(t),
$ with $
   \mat{G}(0) = \mat{I}, \,
$ 
is the fundamental matrix of the system (\ref{eq:indopdesc}).
Evolution of a starting solution is then given by
\begin{equation}
   \label{eq:fundevol}
   \vec{b}(t) = \mat{G}(t) \, \vec{b}(0) .
\end{equation}
For any $T$-periodic $\mat{F}(t)$, there exist matrices $\mat{P}(t)$ and
$\mat{E}$ such that the fundamental matrix can be written
\begin{equation}
   \label{eq:floqthm}
   \mat{G}(t) = \mat{P}(t) \, \ex^{\mat{E}\,t},
\end{equation}
where $\mat{E}$ is independent of $t$ and $\mat{P}(t)$ 
is 
$T$-periodic
(Floquet's theorem).  
It follows immediately that
the change in the solution over one period is given by
\begin{equation}
   \vec{b}(T) = \mat{G}(T) \, \vec{b}(0),
   \qquad
   \mat{G}(T) = \ex^{\mat{E}\,T} = \mat{A} .
\end{equation}
The stability of solutions to
(\ref{eq:indopdesc}) is determined by the eigenvalues $\lambda$
of the constant matrix $\mat{A}$.
If $\vec{b}(0)$ is an eigenvector of $\mat{A}$ 
with eigenvalue $\lambda=\ex^{\sigma_1 T}$,
we find that
$\vec{b}(T+t)=\ex^{\sigma_1 T}\,\vec{b}(t)$ for any $t$.
The real part of the Floquet exponents $\sigma_1$ 
correspond to growth rates of the solutions.  
Although $\mat{A}$ is unknown, 
from (\ref{eq:fundevol}) we see that
the effect of multiplying by $\mat{A}$ 
is equivalent to the result of timestepping through one period.
Therefore we do not have to calculate and store $\mat{A}$ explicitly.
Note that for a steady forcing $\mat{F}$, the period 
$T$ can be chosen arbitrarily.

The eigenvalue problem for $\mat{A}$
is well suited to the Arnoldi process \citep{arnoldi51},
which tends to pick out eigenvalues isolated in the complex plane.  
The many 
decaying modes have $\lambda$ clustered about the origin, marginal 
modes correspond to $|\lambda|$ close to unity.
%
At each iteration we add to the Krylov-subspace given by
$\,\mathrm{span}\{\vec{b}, \mat{A}\vec{b},...,\mat{A}^{k-1}\vec{b}\}$
which we hope contains our solutions. 
In exact arithmetic the $k^{\mathrm{th}}$ Krylov subspace is
equivalent to $\mathrm{span}\{\vec{b}_1,...,\vec{b}_k\}$ 
where the basis vectors $\vec{b}_k$ are obtained from the Arnoldi method.
Numerically the latter set is better suited to span the space.
The Arnoldi process is summarised as follows: (1)
Take a suitable normalised initial basis vector 
$\vec{b}_1=\vec{b}/\|\vec{b}\|_2$.
(2)
At the $k^{\mathrm{th}}$ iteration
evaluate (timestep)
$\tilde{\vec{b}}_{k+1} = \mat{A}\,\vec{b}_k$. 
(3) The result $\tilde{\vec{b}}_{k+1}$ is then
orthogonalised against previous vectors in the basis by the
modified Gram--Schmidt method:
\[
   h_{jk} = \langle \tilde{\vec{b}}_{k+1} , \vec{b}_j \rangle,
   \quad
   \tilde{\vec{b}}_{k+1} := \tilde{\vec{b}}_{k+1} - h_{jk}\vec{b}_j;
   \quad
   j \le k .
\]
(4) Setting $h_{k+1,k} = \| \tilde{\vec{b}}_{k+1} \|_2$, the process 
continues from (2) with the next basis vector 
$\vec{b}_{k+1} = \tilde{\vec{b}}_{k+1} / h_{k+1,k}$.
%
Construct $\mat{Q}_k = \left[ \vec{b}_1,...,\vec{b}_k \right]$
and $\mat{H}_k = \left[ h_{jm} \right ]_{j,m\le k}$.
From steps (3) and (4) we expect $h_{k+1,k}\to 0$.   
In this case,
looking carefully at the steps above, 
the results of the Arnoldi process 
can be written
$\mat{A}\mat{Q}_k = \mat{Q}_k \mat{H}_k$.  
Multiplying on the right by
eigenvectors $\vec{x}$ of $\mat{H}_k$ we find that they are related 
to those of $\mat{A}$ by
$\vec{b} = \mat{Q}_k \vec{x}$.
For non-zero $h_{k+1,k}$ eigenvectors have
residual 
$\|\mat{A}\,\vec{b} - \lambda\vec{b}\|_2 = |h_{k+1,k}| |x_k|$, 
where $x_k$ is the last element of the $k$-vector $\vec{x}$.
Thus, at each iteration eigenvalues $\lambda$ 
of $\mat{H}_k$ are approximate eigenvalues of $\mat{A}$.


In practice the residual $|h_{k+1,k}| |x_k|$
tends to overestimate the error, and in our
calculations the Arnoldi process is stopped when the largest
eigenvalues of $\mat{H}_k$
are sufficiently converged.  The number of iterations required is 
typically of order 100 or less, and so the eigenvalues of the 
small matrix $\mat{H}_k$ can be
cheaply calculated by the QR algorithm.  The memory required to store
the basis vectors scales like $N_h N_r k$.  
It is possible to restart the Arnoldi process with a more suitable
starting vector obtained from the process so far, but without
completely restarting the new process from scratch.  
This implicit restarting
allows further reduction of memory requirements by reducing the
number of basis vectors at each restart \citep{lehoucq98}.  
With $k$ small, restarting
was not found to be necessary, however.
The stencil of the finite
difference scheme does not explicitly affect the storage requirements.
Basis vectors were therefore timestepped with a fourth order finite 
difference scheme.  
Timestepping was performed with the benchmarked code of
Gibbons, Jones and Zhang \citep{christensen01}.

Another advantage of the matrix-free method 
is that, given a timestepping code, only a few extra lines of
code are required to incorporate the Arnoldi process, therefore leaving
significantly less room for error. The code was verified by comparison
with the matrix method used in I--III 
for the steady problem, adjustments for the periodic case in our
matrix-free method are minimal.
Table \ref{tbl:grtest}({\it a}) 
shows the leading two eigenvalues for the
of the steady Kumar--Roberts flow, 
$(D,M)=(0.98354915,0.0001632689)$, 
at $\Rm=1000$ calculated using the method in I--III.
Table \ref{tbl:grtest}({\it b}) shows the same eigenvalues calculated
using our method.  
The higher order radial differences used in the
timestepping code leads to more rapid convergence with $N_r$.
%
Table \ref{tbl:grtest}({\it c}) shows that incorporating the Arnoldi
method accelerates convergence relative to timestepping alone
(and calculation of more than one eigenvalue is possible).
The same starting vector was used for both calculations.
The period $T$ can be chosen arbitrarily for the
steady flow case, but if chosen too small more iterations are required
and therefore more basis vectors must be stored.  For these calculations
we set $T=0.001$.

As the structure of the eigenfunctions varies with $D$ and $M$,
so does the convergence with $N_r$ and $L$.
For most of the following analysis a radial resolution of $N_r=50$ 
and a spectral truncation of $L=18$ was adopted.
Checks at higher resolutions were calculated where growth rates were found
to vary rapidly with the parameters.

\section{Results}
Periodic flows are defined by a closed orbit in $(D,M)$-space.
We restrict ourselves to simple sinusoidal variations in time
with a single frequency $\omega$:
%
\begin{eqnarray}
   D(t)  & = &  D_0 + A_D \cos(\omega t), \nonumber \\
   M(t)  & = &  M_0 + A_M \sin(\omega t).
\end{eqnarray}
It is the aim of this section to assess how the amplitude of variations
$A_{D,M}$ and frequency $\omega$ affect the growth rates and therefore 
the dynamo action.  
\subsection{Magnetic growth rates for time-varying flows}
\begin{figure}
\centerline{\epsfig{file=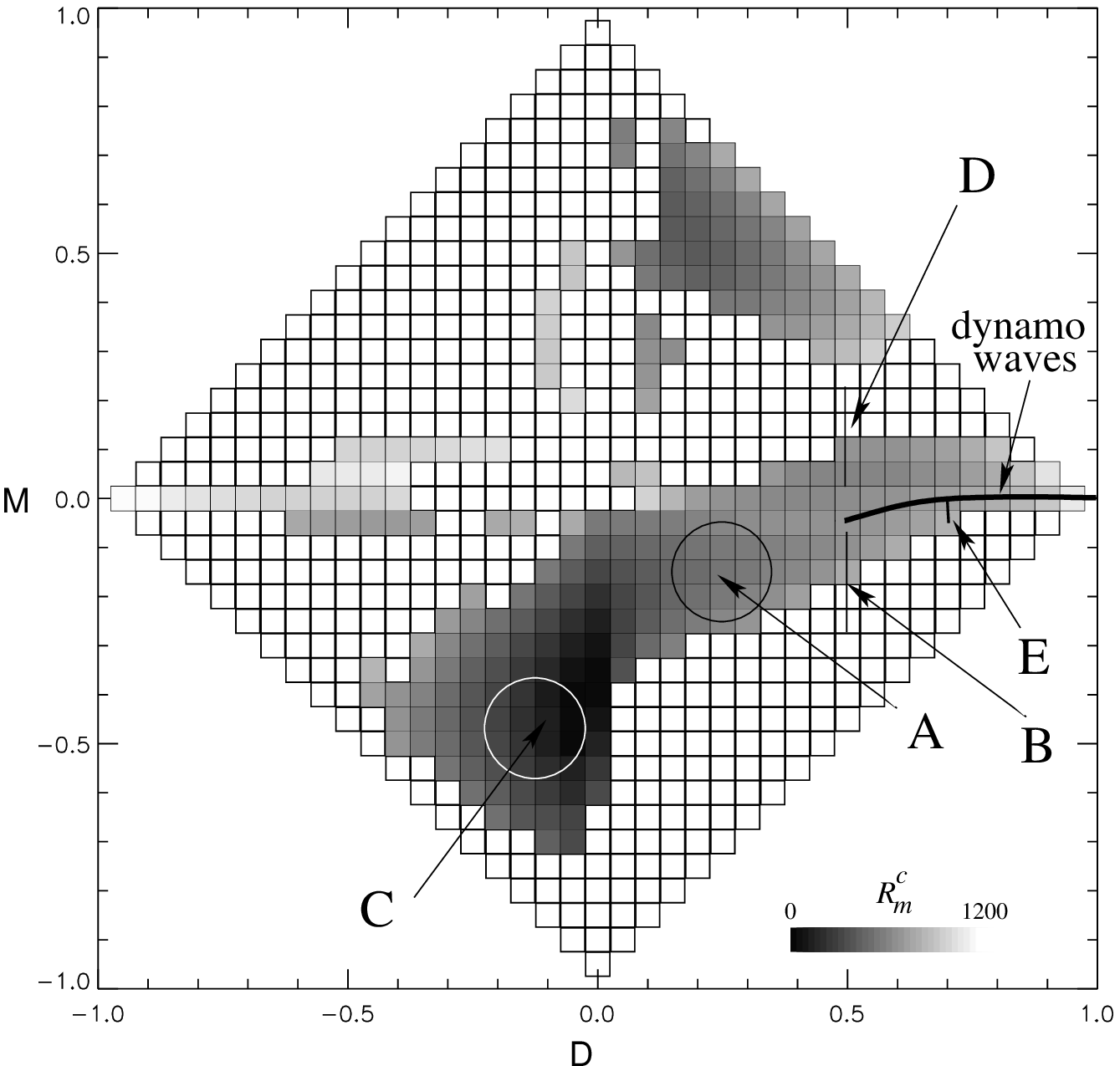,width=12.0cm}}
   \caption{ \label{fig:diamond}
      Regions of dynamo solutions for steady flows and  axial-dipole symmetry for the 
      flow parameterisation: $D$, differential rotation; $M$ 
      meridional circulation. The thick line running into the right
hand apex contains oscillatory solutions related to the dynamo waves of
the Braginsky limit. Other lines and circles describe periodic flows mentioned
in the text.
   }
\end{figure}

\begin{figure}
   \epsfig{figure=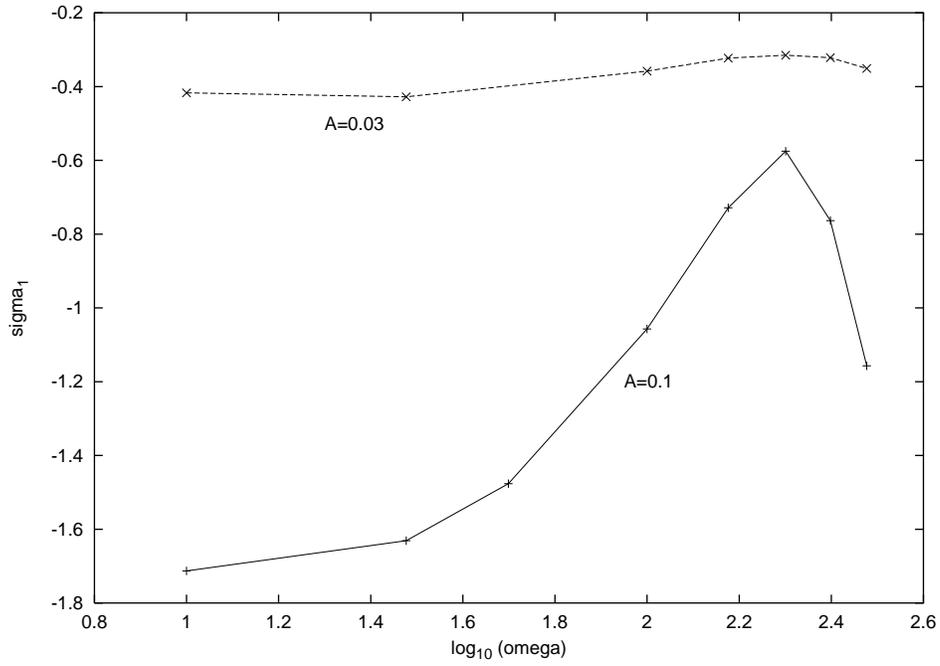}
   \caption{ \label{fig:grAamp}
      Growth rates $\sigma_1$ {\it vs}. $\omega$ for different amplitudes 
      $A_D=A_M=A$ about the point $(0.25,-0.14)$.
   }
\end{figure}

Figure \ref{fig:grAamp} shows growth rates for different amplitudes of
variations about the point $(D_0,M_0)=(0.25,-0.14)$, 
marked A in Fig.~\ref{fig:diamond}, which lies on a 
line of minimum $\Rm$ extending from the Braginsky limit point $(1,0)$ 
(see I, Table~5). The majority of neighbouring points 
have lower growth rates for the given value of $\Rm=87$. 
Figure \ref{fig:grAamp} shows that the effect of oscillations of the
flow on the growth rates is more pronounced with increased
oscillation amplitude.
%

\begin{figure}
   \epsfig{figure=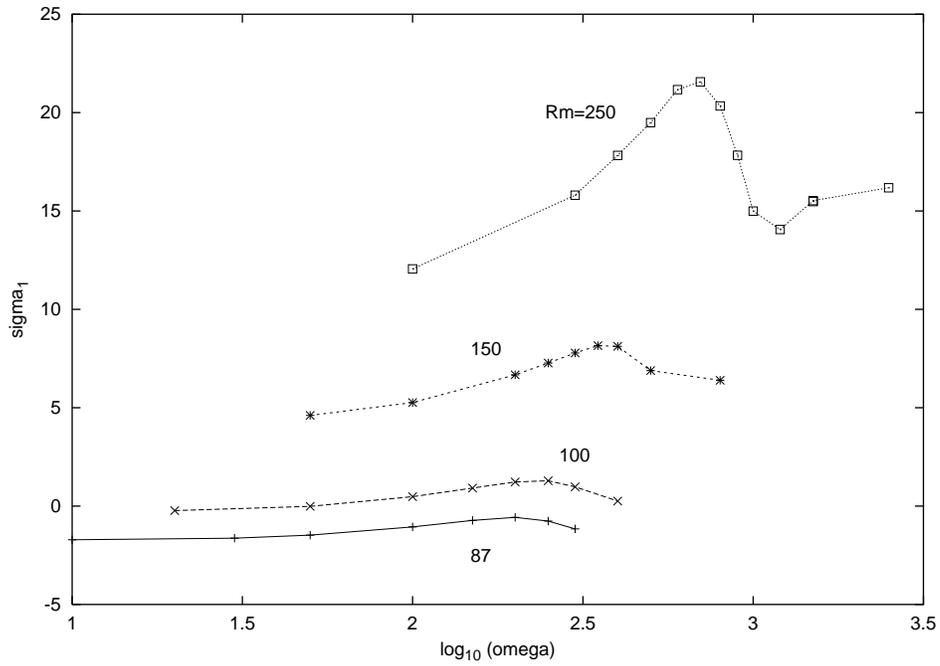}
   \caption{ \label{fig:grARm}
      As Fig.\ \ref{fig:grAamp} for various $\Rm$; $A_D=A_M=0.1$,
      A in Fig.\ \ref{fig:diamond}.
   }
\end{figure}

For a steady flow, given any small real number $\varepsilon>0$ 
there exists a finite
time $t$ such that transients are reduced so that 
$|\sigma-\sigma(t)|<\varepsilon$, where $\sigma$ is the growth
rate corresponding to the steady flow at each point on the orbit,
and $\sigma(t)$ is the growth rate of an 
arbitrary initial field as it adjusts to the flow.
Provided that growth rates are
piecewise continuous (for example $D(t)$ could be discontinuous but
periodic, see Backus 1958), 
a frequency $\omega$ can always be selected low
enough such that net growth rate over the cycle is close to
the mean $\bar{\sigma}$ of those on the orbit.
The limit $\sigma_1(\omega)\to\bar{\sigma}$ as $\omega\to 0$
is observed in the numerical 
calculations. What is more interesting, however, is that with 
finite $\omega$ the dynamo can do much better than this mean,
$\sigma_1(\omega)>\bar{\sigma}$,
as seen in most figures for the growth rate.
Figure \ref{fig:grARm} shows that the effect increases
with $\Rm$ and that the peak occurs at a frequency $\omega$
that increases in proportion to $\Rm$.

\begin{figure}
   \begin{tabular}{cccc}
      & $\omega\to 0$ & $\log\omega=2.6$ & $\log\omega=2.9$ \\
      $\frac{8}{16}T$ &
      \epsfig{figure=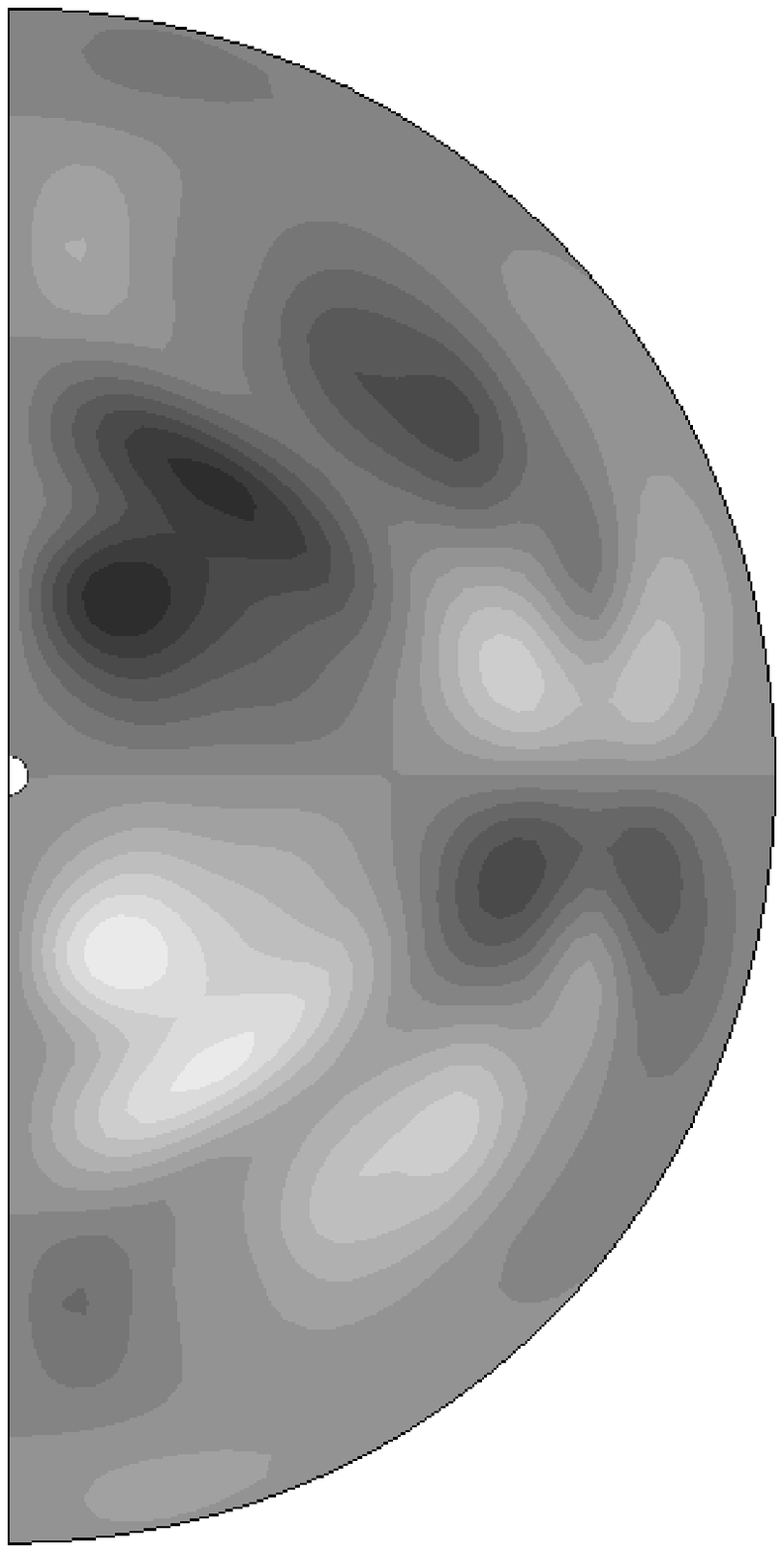, scale=0.35} &
      \epsfig{figure=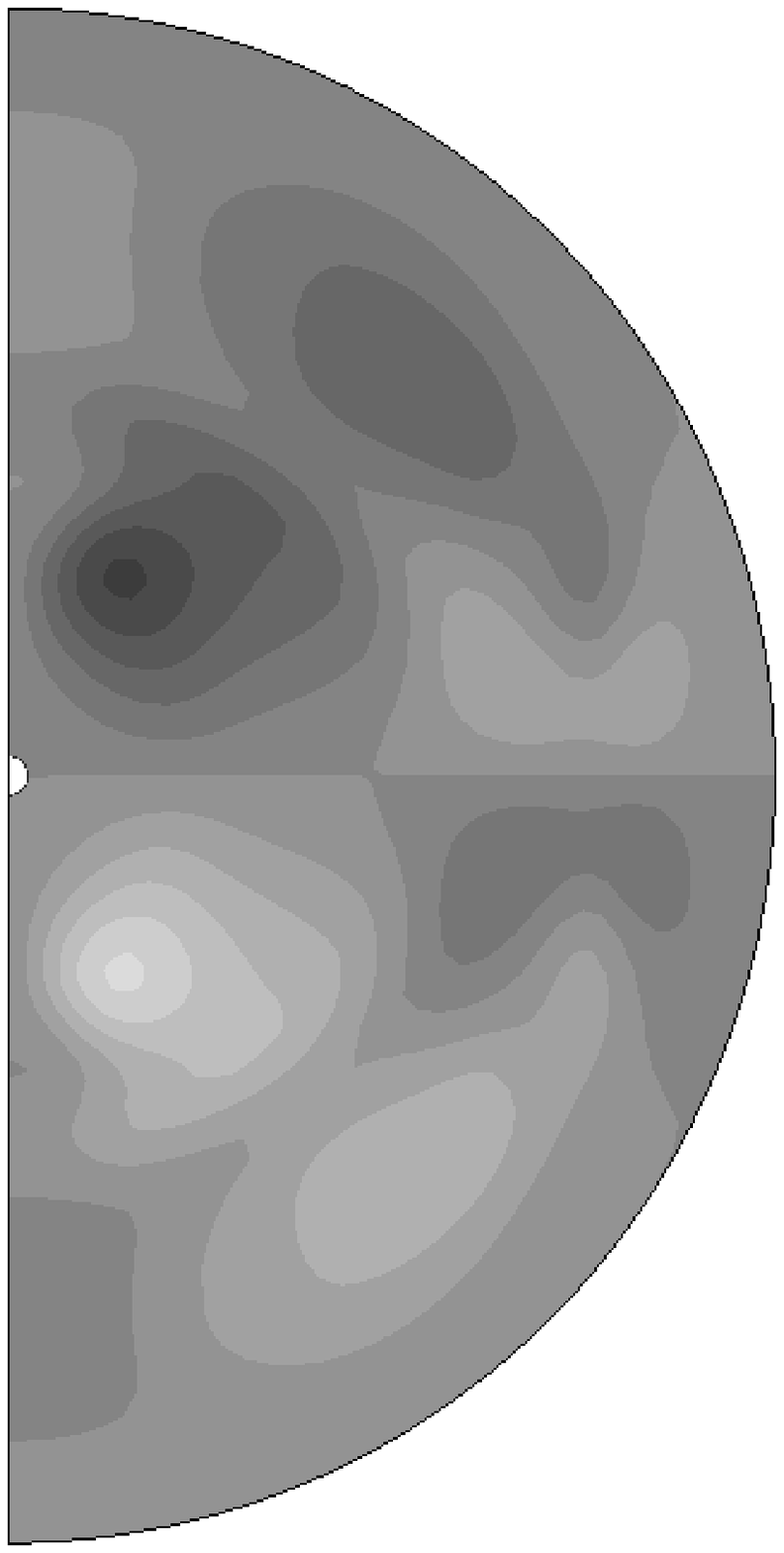, scale=0.35} &
      \epsfig{figure=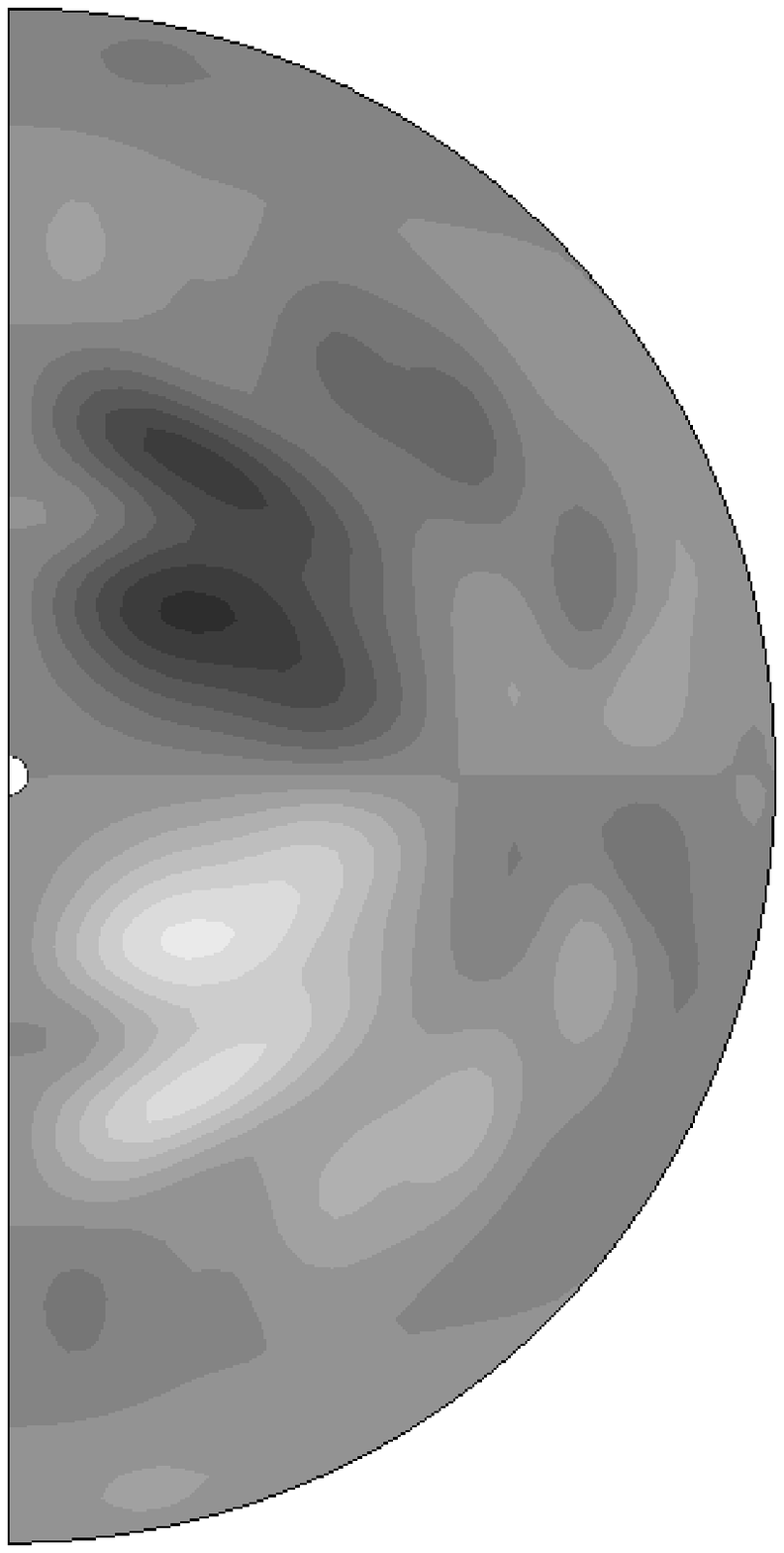, scale=0.35} \\[8pt]
      $\frac{13}{16}T$ &
      \epsfig{figure=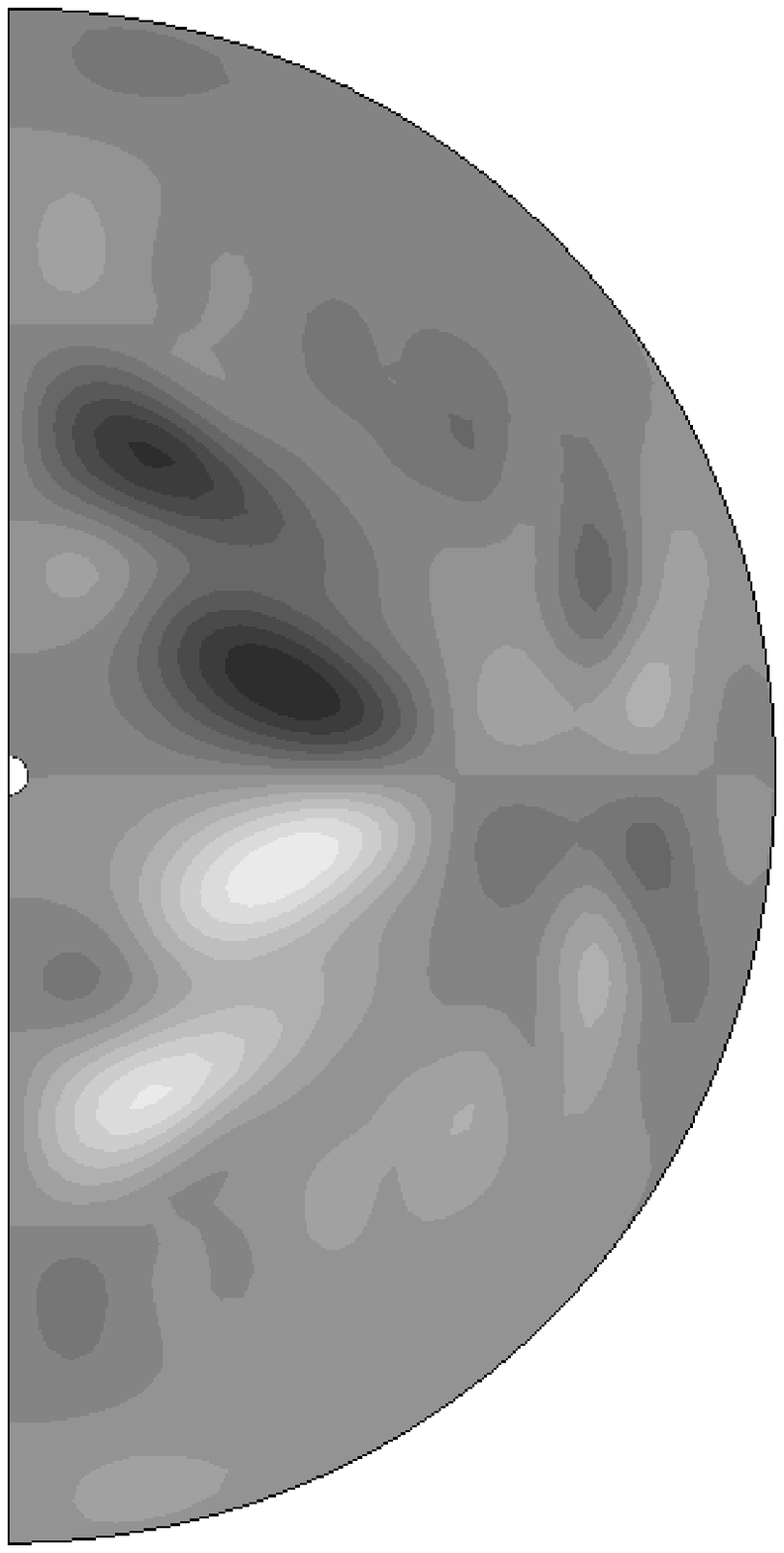, scale=0.35} &
      \epsfig{figure=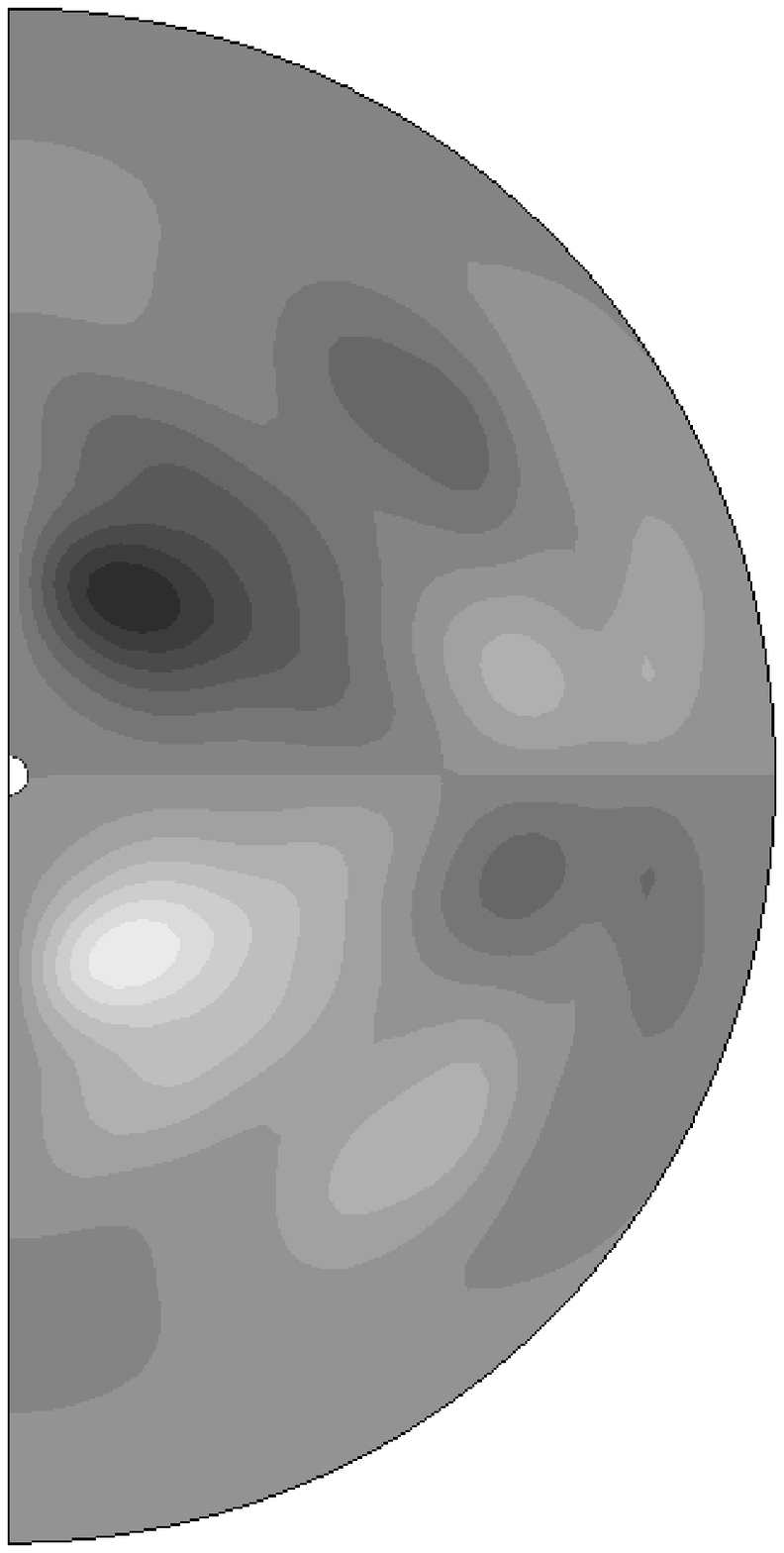, scale=0.35} &
      \epsfig{figure=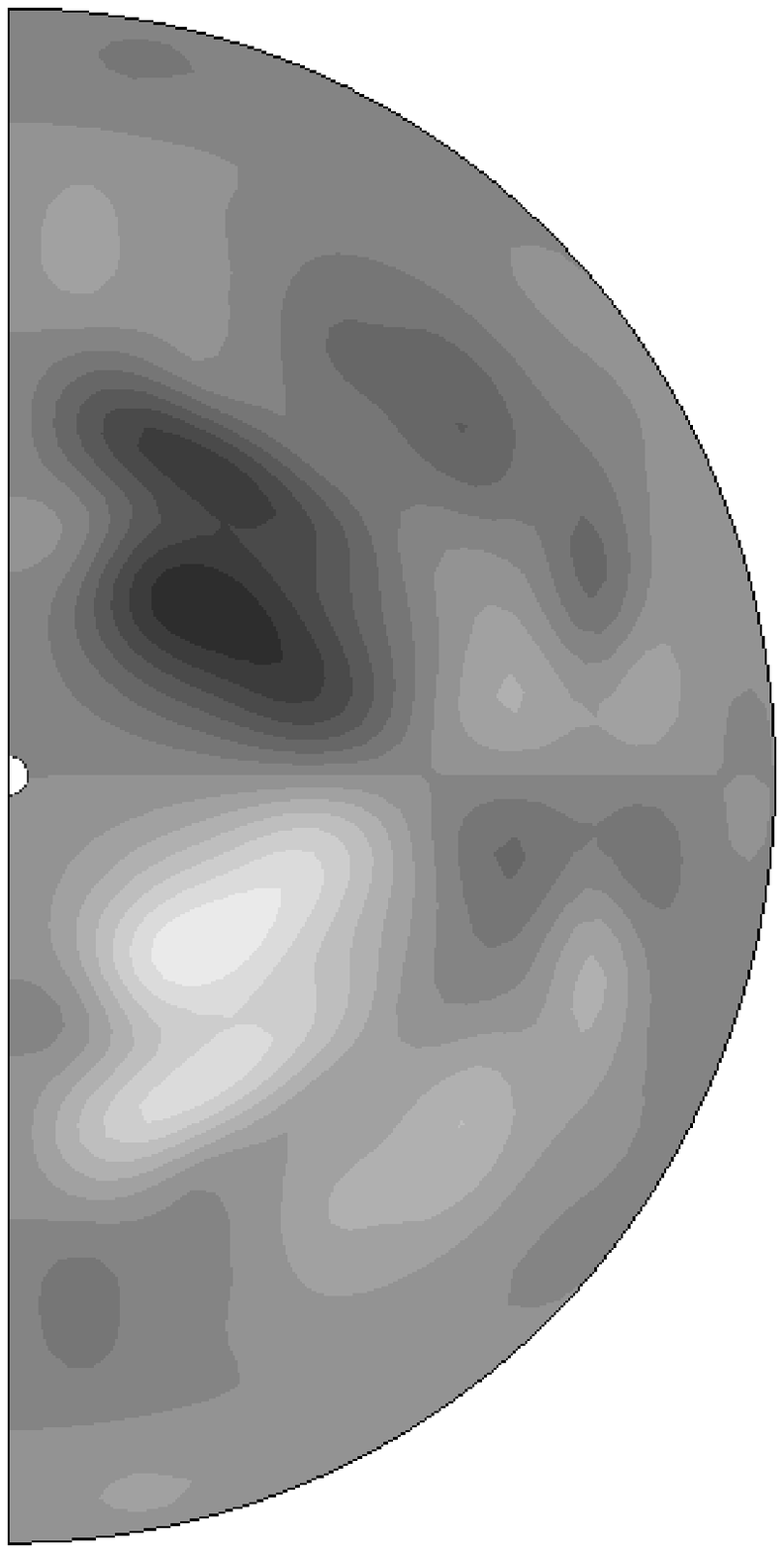, scale=0.35} 
   \end{tabular}
   \caption{ \label{fig:AmerBph90}
      Meridional sections, $B_\phi$, $\phi=\pi/2$; 
      $\Rm=150$, $(D,M)=(0.25,-0.14)$, $A_D=A_M=0.1$.  
      Eigenfunctions were calculated for steady flows at the appropriate 
      point and are plotted with independent contour values; 
      for $\omega\ne 0$, contours are plotted for the
      same values at different times.  
      At $\log\omega=2.6$ 
      the field structure looks smoothed. At $\log\omega=2.9$ 
      the flow changes too quickly for the field to respond and 
      the field is almost steady.
   }
\end{figure}

Meridional sections of the magnetic field for this flow are plotted in 
Fig.\ \ref{fig:AmerBph90}.  The two times correspond closely to the
points on the orbit which have the maximum (upper row) and 
minimum growth rates for steady flows ($\omega\to 0$).  
The structure of these eigenfunctions is similar;
regions of $B_\phi$ are generally well separated in space.
The dissipation for these fields 
is larger for the lower panel where fields of opposite 
sign are squeezed towards the equator.
For non-zero $\omega$ the location of the flux changes over
the cycle, and at $\log\omega=2.6$ the field represents a smoothed version
of the two eigenfunctions.  Fewer small-scale features are present and
the flow performs well as a dynamo (see Fig. \ref{fig:grARm}).
At $\Rm=150$ the peak frequency for $\sigma_1$ 
is $\log\omega\approx 2.6$; 
taking $T\sim (\delta/d)^2$ 
as an approximate timescale for diffusion,
if $\delta\sim d/8$ is an approximate length scale for the small
scale features of the eigenfunctions, 
we find that the timescale for diffusion and for the
peak flow oscillation coincide.
The magnetic field is then smoothed effectively.
Above this frequency the growth rate decreases again as the field is 
unable to respond to rapid changes in the flow.  Spatial smoothing is
lost and the field is close to steady --- plots at the two times
for $\log\omega=2.9$ are almost identical.
The field responds as though to a steady flow, retaining the 
stronger (smaller-scale) features from each eigenfunction.

\begin{figure}
   \epsfig{figure=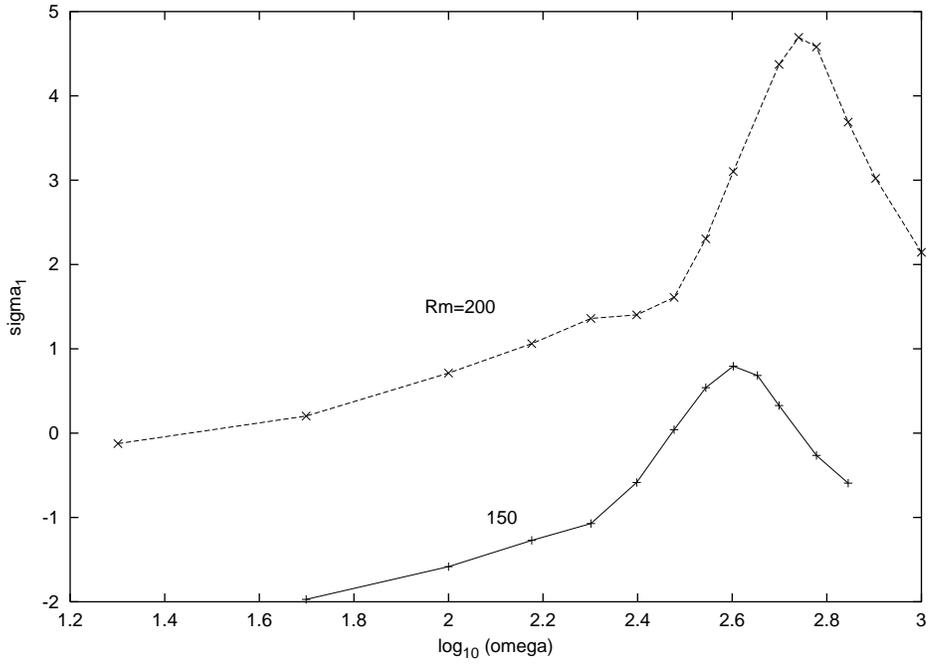}
   \caption{ \label{fig:grBRm}
      Growth rates $\sigma_1$ {\it vs}.\ 
      $\omega$ for an orbit about the point 
      $(0.5,-0.15)$; $A_D=0$, $A_M=0.1$ (Fig.\ \ref{fig:diamond}, B).
   }
\end{figure}

Figure \ref{fig:grBRm} shows growth rates for an orbit about the point
$(0.5,-0.15)$, marked B in Fig.~\ref{fig:diamond}, which   
lies on the lower boundary of the region of successful steady dynamos.
$D$ remains constant and $M$ varies to carry the flow outside the
dynamo region.  The time-dependent flow produces a positive effect on
the growth rate.  The spatial structure of the eigenvectors on this orbit is
similar to that of the previous point considered, with well separated regions
of positive and negative azimuthal field. For $R_m=150$ there is dynamo
action only for frequencies $\log\omega\approx 2.6$, and the average 
growth rate around the orbit is negative.

\begin{figure}
   \epsfig{figure=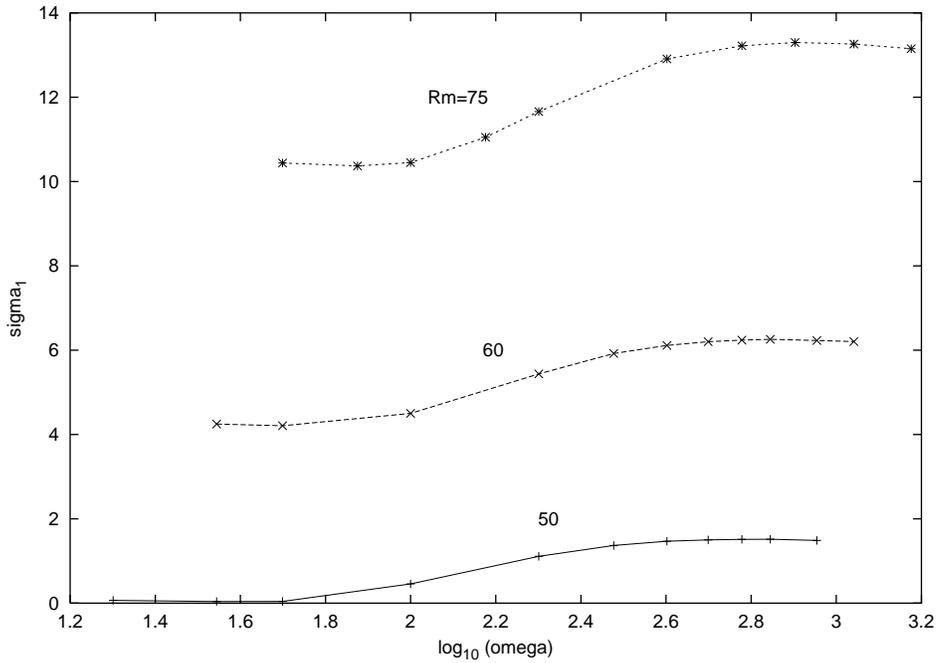}
   \caption{ \label{fig:grGRm}
      Rising growth rates for an orbit about $(-0.1,-0.45)$; 
      $A_D=A_M=0.1$  (Fig.\ \ref{fig:diamond}, C).
   }
\end{figure}

We now describe a case where meridional circulation is greater than
differential rotation, $(D,M)=(-0.10,-0.45)$, 
marked C in Fig.~\ref{fig:diamond}.  
This point is close to where the critical magnetic 
Reynolds number for steady flows is at a global minimum.  
Once again time dependence of the flow enhances the growth rate,
Fig.~\ref{fig:grGRm}.  A small rise can be seen in the
growth rate, although less significant relative to the increase 
associated with an increase in the magnetic Reynolds number.  
Being close to the point where $\Rm^c$ is a minimum,
the fields are already relatively large scale and the smoothing 
effect of oscillations therefore has a smaller effect on the growth rate.

\begin{figure}
   \epsfig{figure=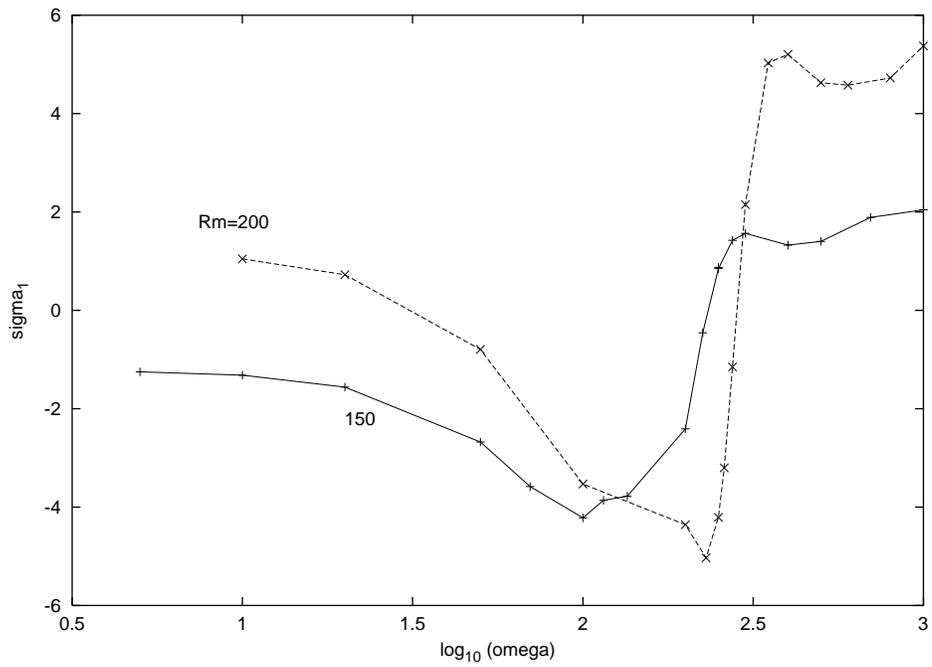}
   \caption{ \label{fig:grCRm}
      Growth rates for increasing $\omega$ about the point $(0.5,0.11)$; 
      $A_D=0$, $A_M=0.1$  (Fig.\ \ref{fig:diamond}, D).
   }
\end{figure}

\begin{figure}
   \begin{tabular}{cccc}
      & $\omega\to 0$ & $\log\omega=2.13$ & $\log\omega=2.60$ \\
      $\frac{4}{16}T$ &      
      \epsfig{figure=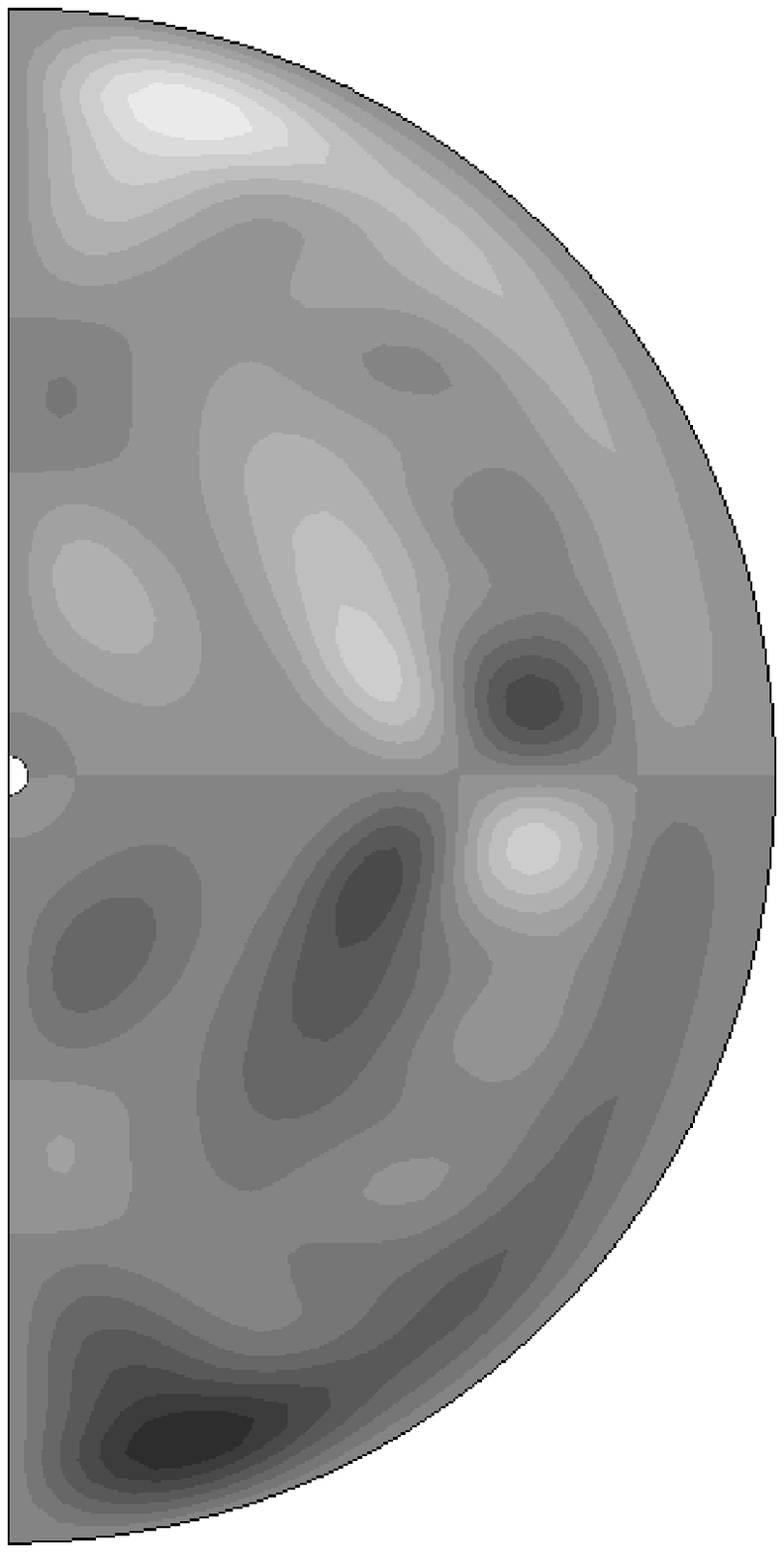, scale=0.35} &
      \epsfig{figure=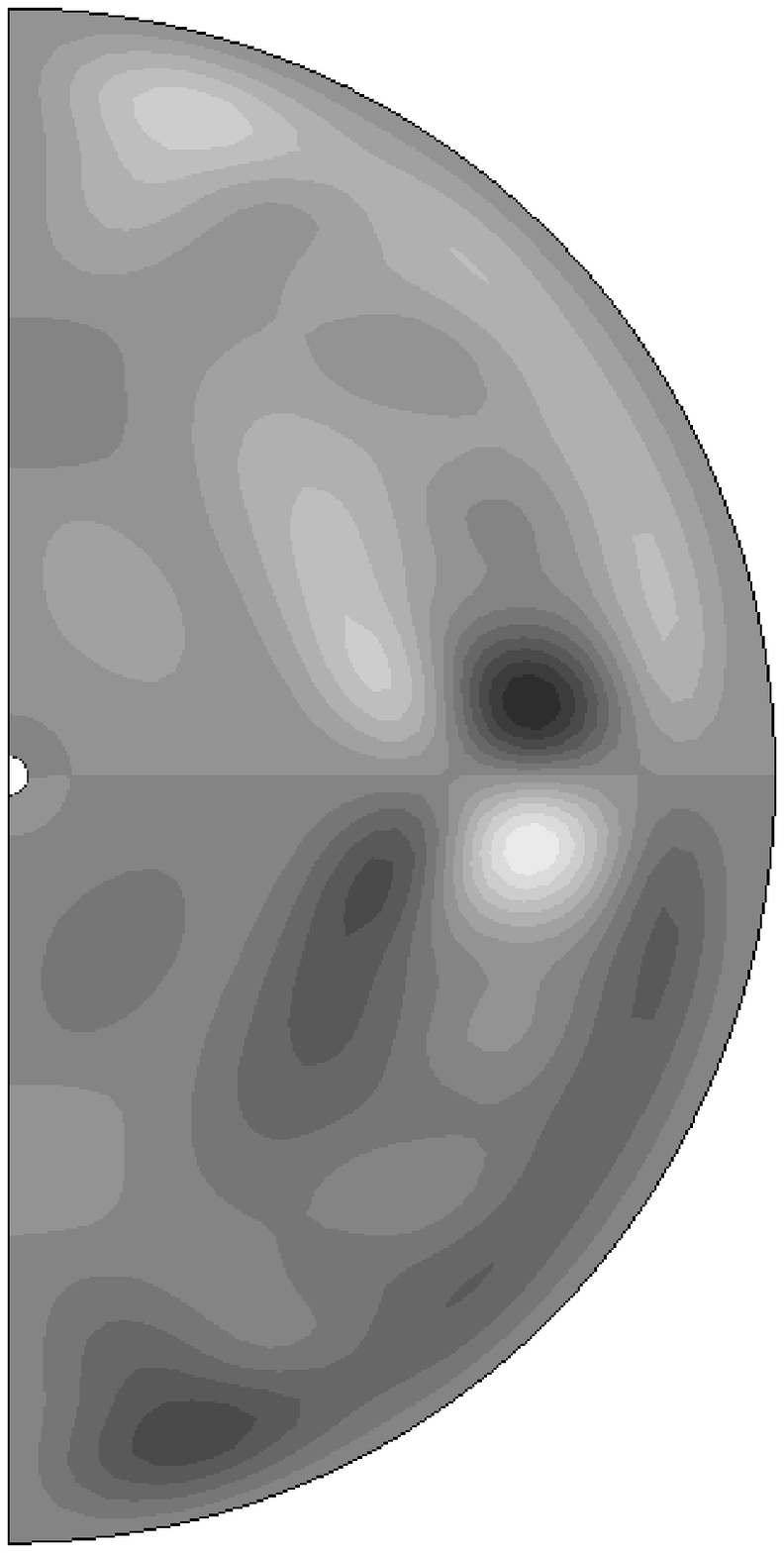, scale=0.35} &
      \epsfig{figure=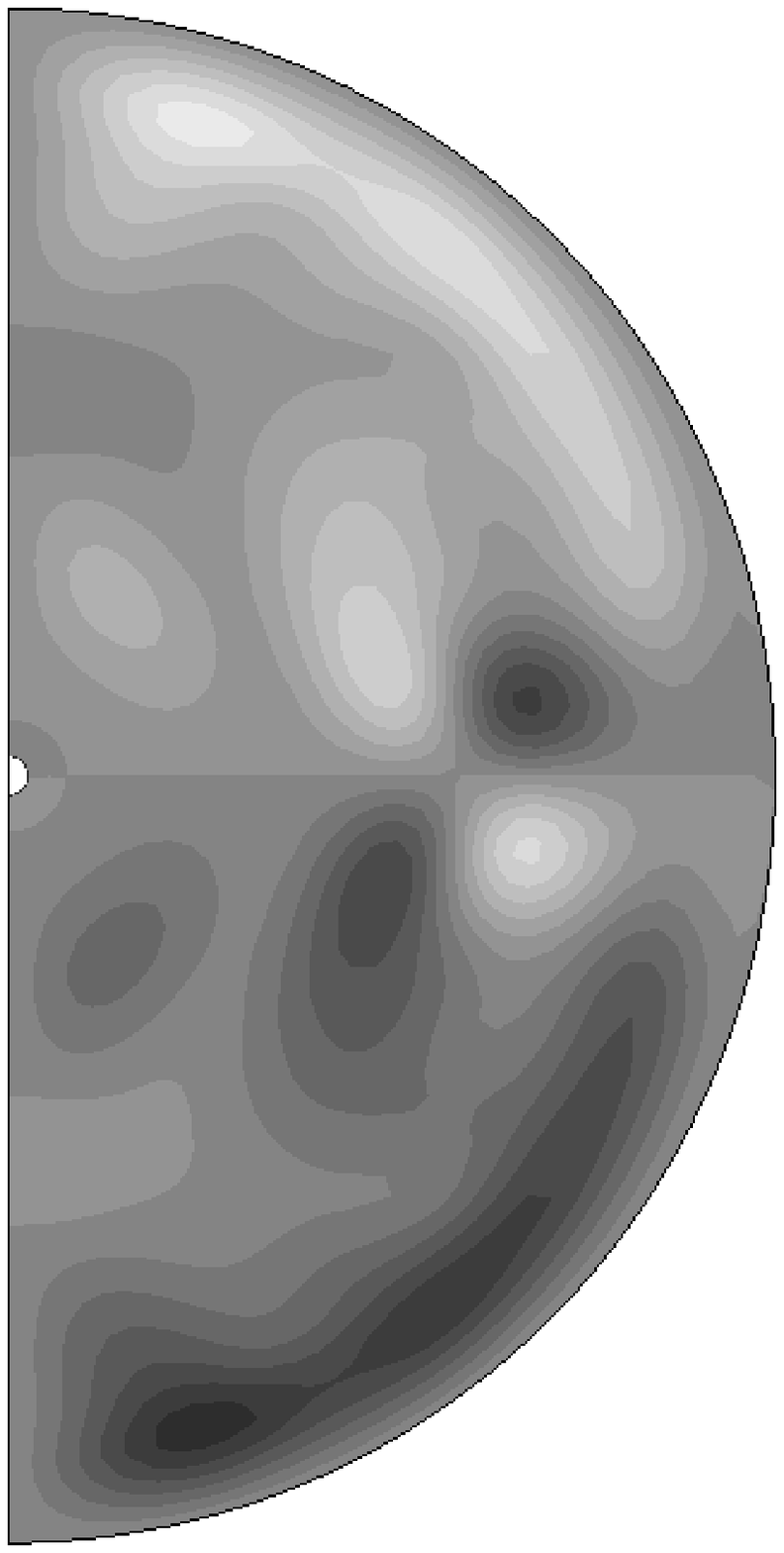, scale=0.35} \\[8pt]
      $\frac{12}{16}T$ &
      \epsfig{figure=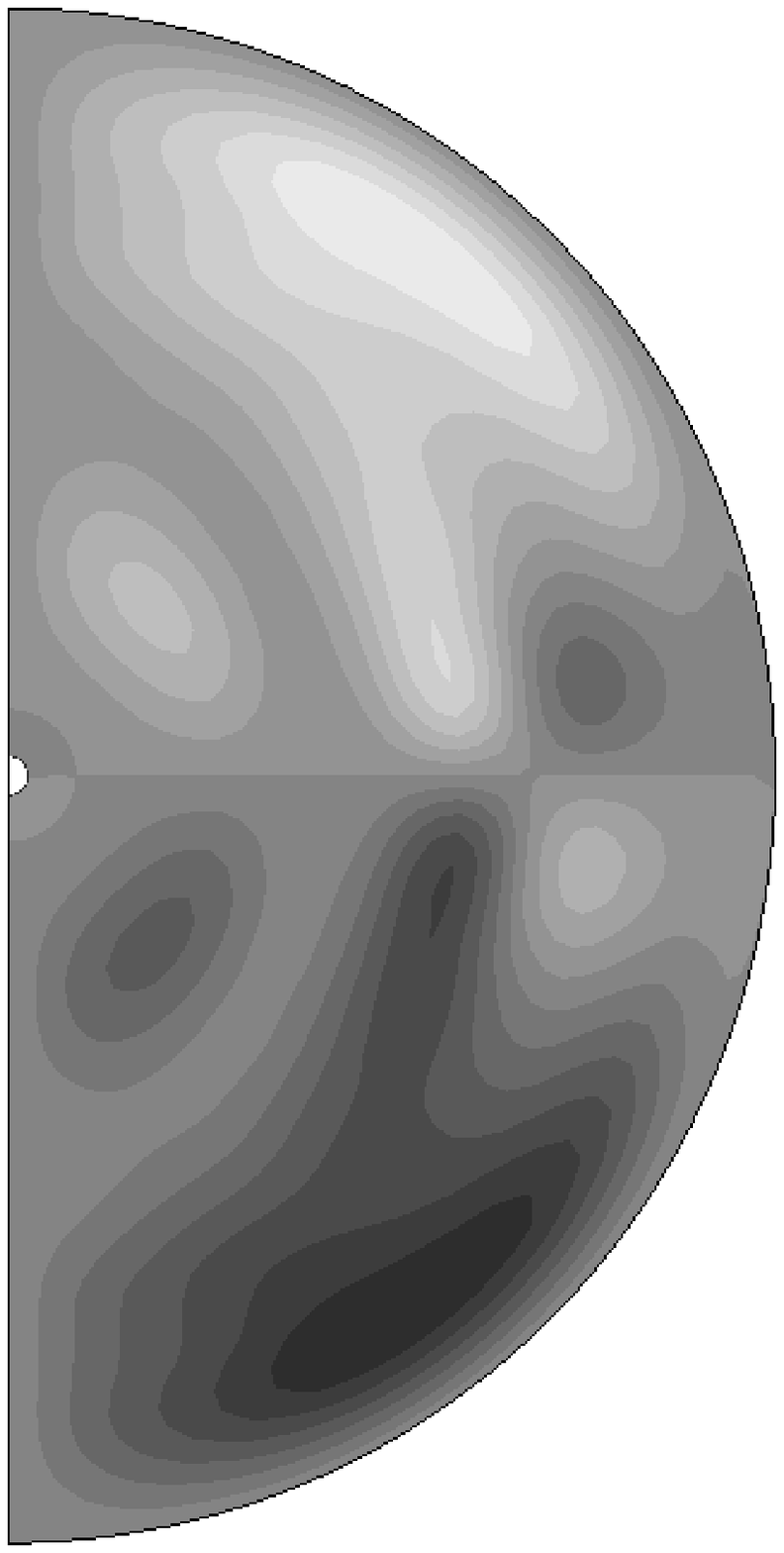, scale=0.35} &
      \epsfig{figure=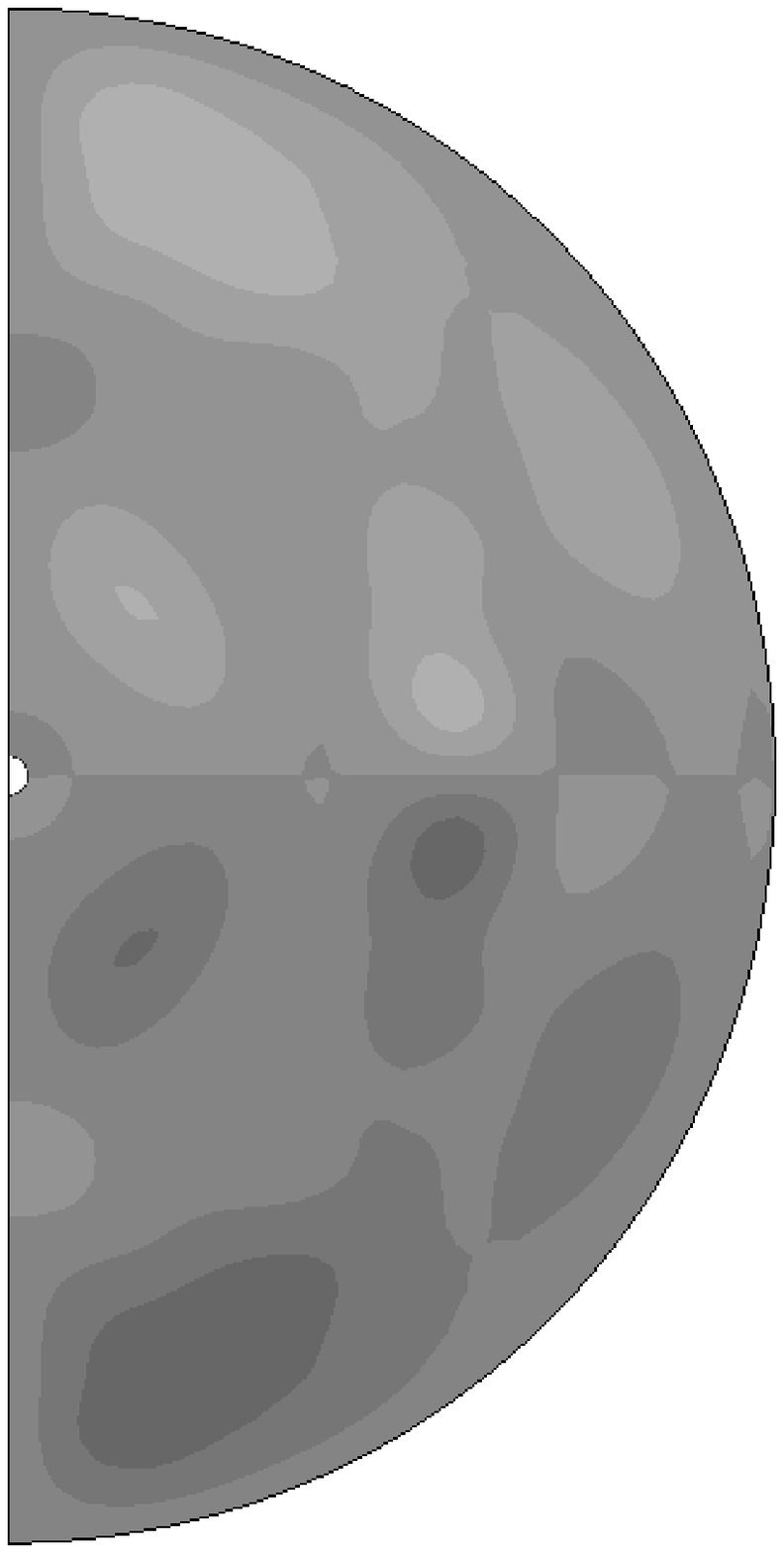, scale=0.35} &
      \epsfig{figure=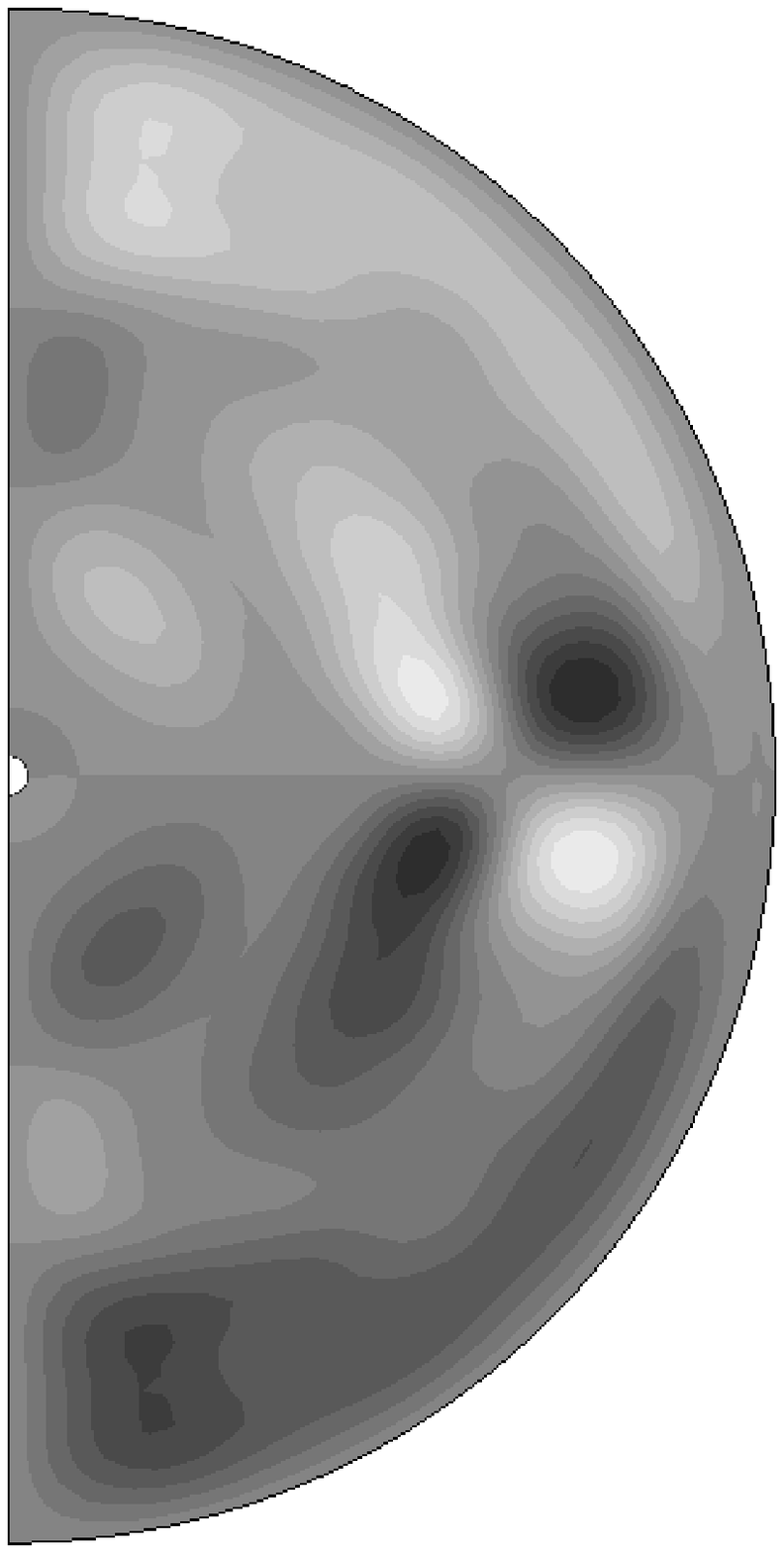, scale=0.35}
   \end{tabular}
   \caption{ \label{fig:CmerBph00}
      Meridional sections, $B_\phi$, $\phi=0$; 
      $\Rm=150$, $(D,M)=(0.5,0.11)$, $A_D=0$, $A_M=0.1$.  
      The times $t=\frac{4}{16}\,T$ and $\frac{12}{16}\,T$ correspond to
      maximum and minimum $M(t)$ respectively.  There is a radial
      shift of the `clover' pattern near the equator for the eigenfunctions
      ($\omega\to 0$).  
      Closely proximity of opposing flux leads to cancellation seen 
      towards the outer boundary, lower panel with $\log\omega=2.13$.  
      At $\log\omega=2.60$ the field is more steady.
   }
\end{figure}

Growth of the time-dependent solutions is not always found to be
better than the mean of the growth rates on the orbit.  
Figure \ref{fig:grCRm} shows the case $(D,M)=(0.50,0.11)$, 
marked D in Fig.~\ref{fig:diamond},
where oscillations in the flow are initially 
damaging to the dynamo.
If the oscillations are sufficiently rapid, however, the dynamo is 
again able to perform better.  Meridional sections are shown in 
Fig.\ \ref{fig:CmerBph00}.  
The regions of strongest flux are located very 
close together on the equator.
Over the cycle radial shifts of the clover-leaf pattern of flux causes
considerable overlap of opposite signs.  This occurs mostly 
towards the outer edge of the equatorial region.
As the flow oscillates, in the lower plot for
$\log\omega=2.13$ it can be seen that there is cancellation of flux in the
outer region.  Some flux remains at the other regions where the signs
for the two eigenfunctions do correlate.  This cancellation of fluxes
over the cycle leads to reduced growth rates.  When the oscillation is
much faster, however, the dynamo does not have time to generate
flux of opposing sign.  The field is more steady for $\log\omega=2.60$ 
and has a larger growth rate.

The radial field in the above has little structure of interest.
It is concentrated mainly on the axis with opposite sign in each
hemisphere (see I, Fig. 9{\it a,b}).  
The structure does not change appreciably over the cycles.
In strength, however, it is observed to wax and wane.

%
%
%

\subsection{Dynamo wave solutions}

Meridional circulation has been seen to play a key part in reversals.  
\citet{sarson99} have studied a system in which irregular reversals
are linked to a drop in meridional circulation, leading to a preference
for oscillatory fields.  
\citet{wicht04} have recently studied a reversal mechanism that involves 
an advection of reversed flux by a large-scale S1 flow.  
Reversals occur quasi-periodically in their model.
This behaviour may be related to the dynamo wave solutions obtained in
III. The oscillation has the form of a dynamo wave in which flux migrates
along the longitudes defined by the downwellings of the
convective parts of the flow, which could partially explain the observed
tendency for virtual geomagnetic poles to track around the Pacific during
polarity transition \citep{gubbins94}. The steady flow model can only oscillate
periodically, but we can construct a more geophysically realistic reversal by choosing a 
time-dependent flow that traces an orbit in $(D,M)$-space that strays into the
dynamo wave region, depicted by the line E in Fig.~\ref{fig:diamond}, for
a fraction $f$ of its period. The field behaviour will depend on the frequency of the
dynamo wave, $\omega_D$ and the time spent by the flow in the oscillatory regime.
If $\omega\gg f\omega_D$ the flow will only spend a brief time in the oscillatory 
regime and we expect only a minor change in the magnetic field. If $\omega\ll f\omega_D$
the flow spends a long time in the oscillatory regime and we expect the field
to oscillate several times before becoming steady again. The interesting
case is when $\omega\approx f\omega_D$, when the field may only have time
to oscillate for one or a half cycle, producing an excursion or a reversal 
respectively.

We now explore reversal behaviour using periodic flows. 
The structure of the eigenfunctions for steady flows changes appreciably as
$M$ crosses zero.  It was seen in the previous section that, where this
is the case, fluctuations are not necessarily good for the dynamo.
Instead, an orbit is chosen to enter the oscillatory range from the 
negative side.  Consider the flow defined
by fixed $D=0.7$ and $M$ varying sinusoidally between $-0.0140$ and $-0.0020$
(E in Fig.~\ref{fig:diamond}).
This orbit spends approximately one third of the time within the band of 
oscillatory solutions reported in III, 
which lies between $M=-0.0057$ and $-0.0010$.
The dynamo wave frequency for steady flows increases 
with $R_m$ and appears to saturate 
at about $\Im(\sigma)=17$ (see III, Figs~2,3); 
it appears to be limited by the diffusion time. 
Here, $\omega$ for the time-dependent flow must be chosen comparable
with this frequency to give a single reversal, a value which is too low to 
assist the dynamo action significantly. 



\begin{figure}
   \epsfig{figure=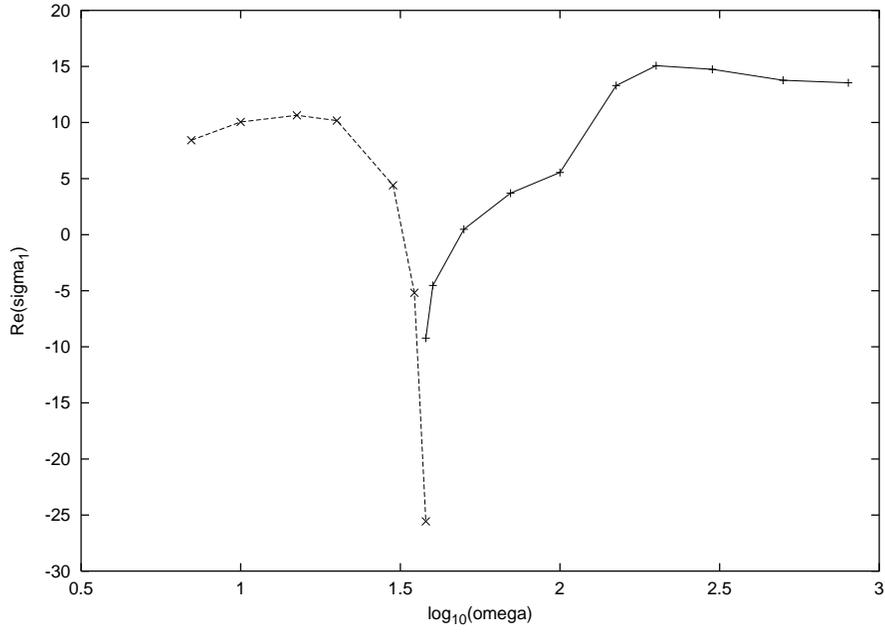, scale=0.95}
   \caption{ \label{fig:grDosc}
      $\Re(\sigma)$, $\Rm=700$ at 
      $(D,M)=(0.7,-0.008)$, $A_D=0$, $A_M=0.006$ 
      (Fig.\ \ref{fig:diamond}, E).  For the dashed curve
      $\Im(\sigma_1)=\omega/2$.
   }
\end{figure}
\begin{figure}
   \epsfig{figure=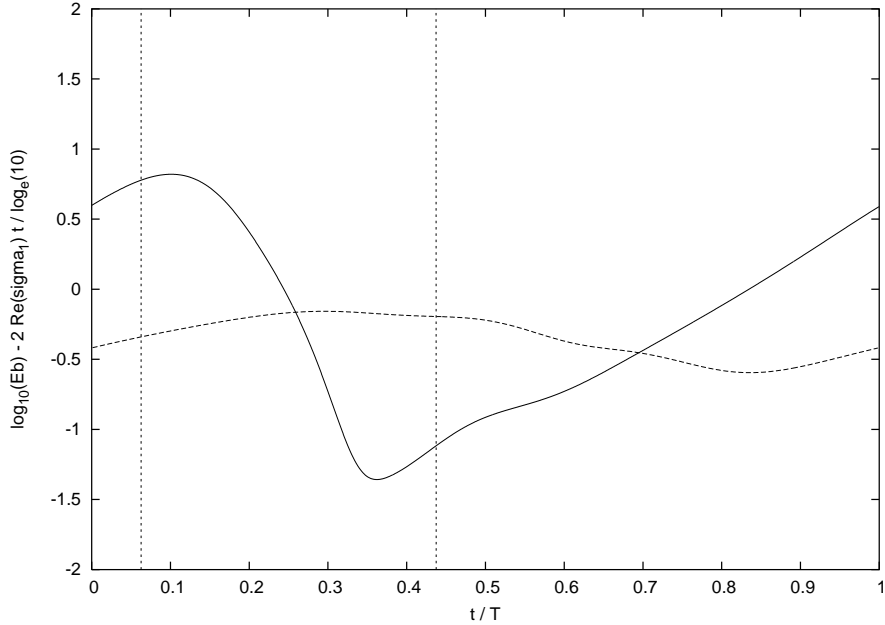, scale=0.95}
   \caption{ \label{fig:D20mnrg}
      Magnetic energy (minus net growth) for parameters as
      Fig.\ \ref{fig:grDosc}. Reversing solution $\log\omega=1.30$ (solid);
      failed reversal $\log\omega=2.00$ (dashed).
      Vertical bars represent the period in the oscillatory regime.
   }
\end{figure}

Figure \ref{fig:grDosc} shows the the growth rate for the time dependent
flows as a function of frequency $\omega$. 
At $\Rm=700$, $\Im(\sigma)$ is approximately $10$ 
for steady flows in the oscillatory range. 
Reversing solutions may be expected for 
$Tf\gtrsim\pi/\Im(\sigma)$, 
or equivalently $\omega\lesssim 20/3$, $\log\omega\lesssim 0.8$
as the time in the oscillatory range is approximately one third of the
cycle. 
Reversing solutions (dashed curve) are observed for $\omega$
larger than the dynamo wave frequency, although
growth rates fall quickly when the
period of the flow oscillation is too short to be 
compatible with the period of the
oscillatory solution.  If the time within the oscillatory region
leads to only a half-complete reversal, the field exiting the 
region bears little resemblance to the entering field, 
which is much like the eigenfunction for these low $\omega$,
and therefore leads to reduced growth rates.
For greater $\omega$
the field does not spend sufficient time within the
oscillatory region to reverse (solid curve) and at higher $\omega$
the growth rates are increased.
The magnetic energy for a typical reversing solution is plotted
in Fig.\ \ref{fig:D20mnrg},
showing a 
drop as the solution passes through the 
oscillatory region where the reversal occurs.  
It is possible to vary $M$ so that $f$ is less than a third.  However,
for the reversing solution in Fig.\ \ref{fig:D20mnrg},
the smooth growth rate curve,
while outside the oscillatory region, indicates the 
field quickly becomes independent of the period within the
oscillatory region, apart from in sign, due to the slow period of 
the flow oscillation.

\begin{figure}
   \begin{tabular}{cc}
      $0.05\,T$
      \epsfig{figure=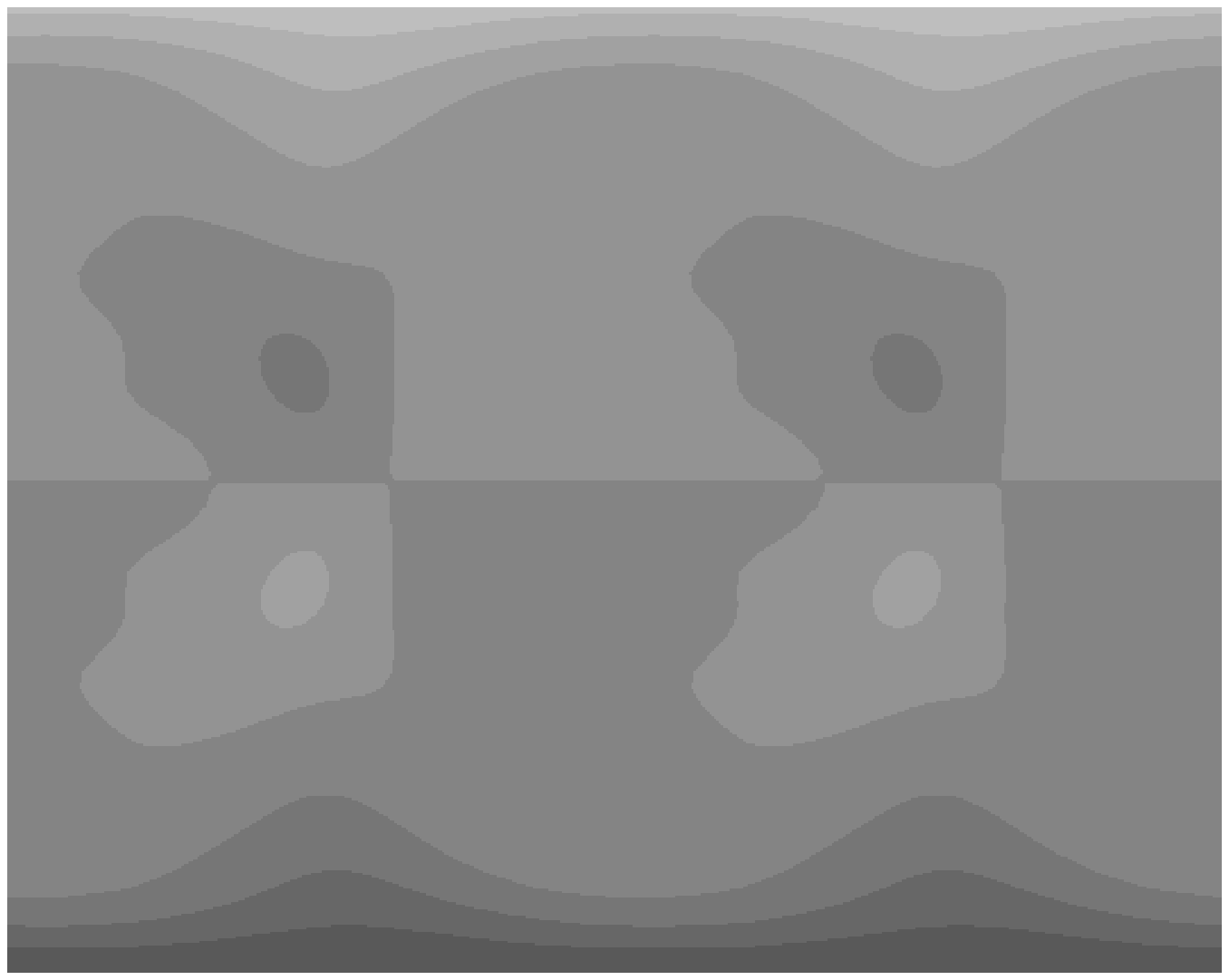, scale=0.35} &
      \epsfig{figure=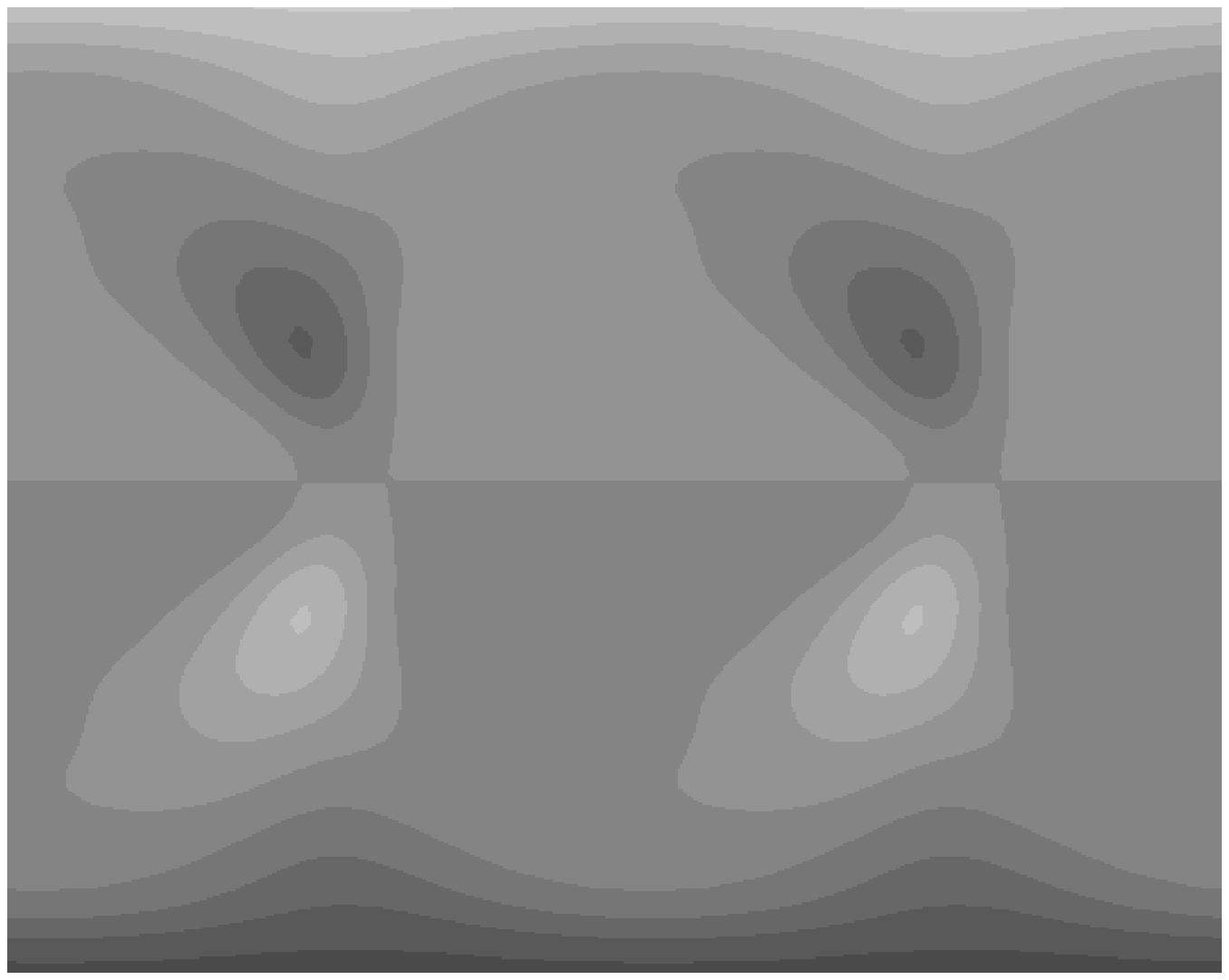, scale=0.35} 
      $0.15\,T$ \\
      $0.25\,T$
      \epsfig{figure=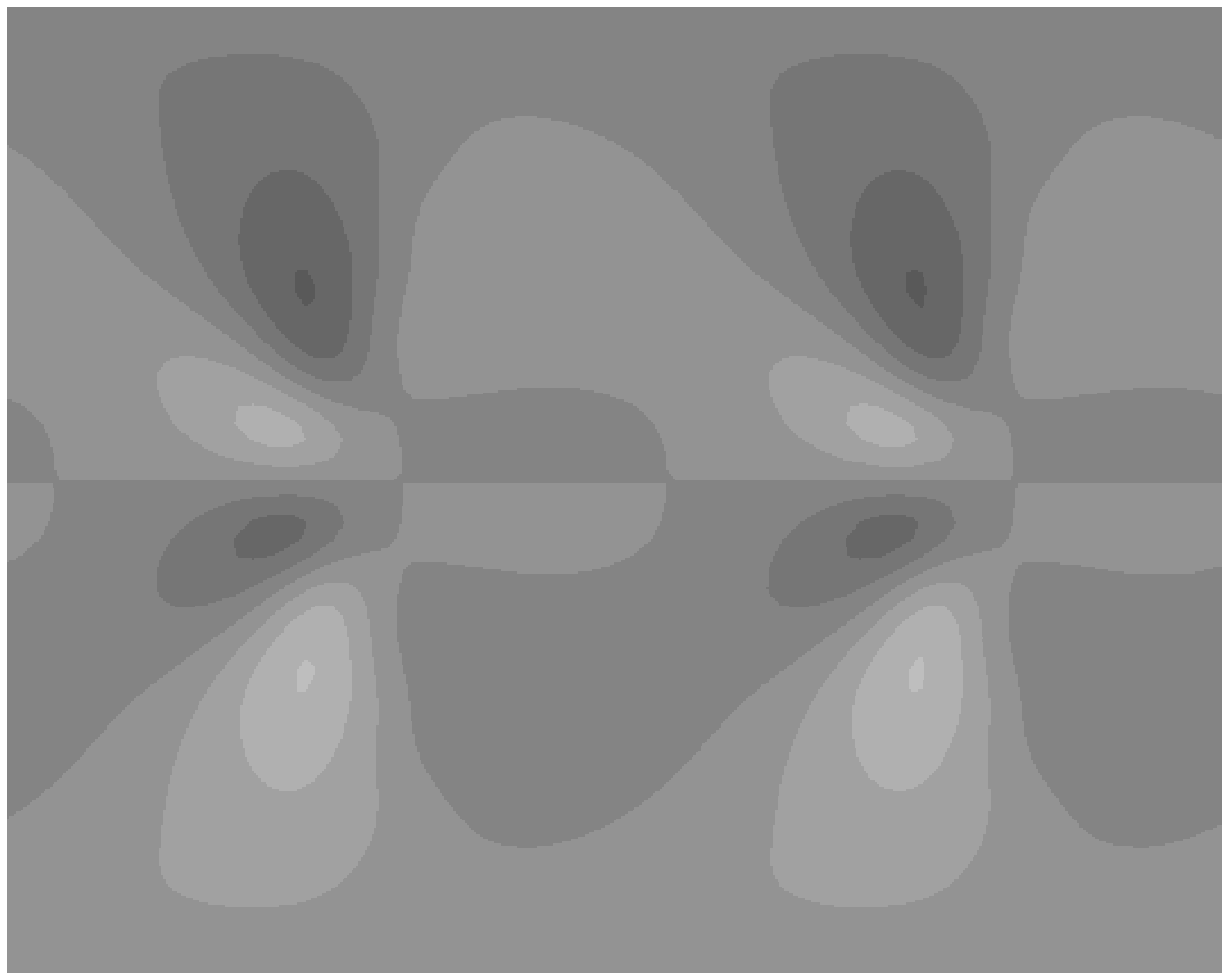, scale=0.35} &
      \epsfig{figure=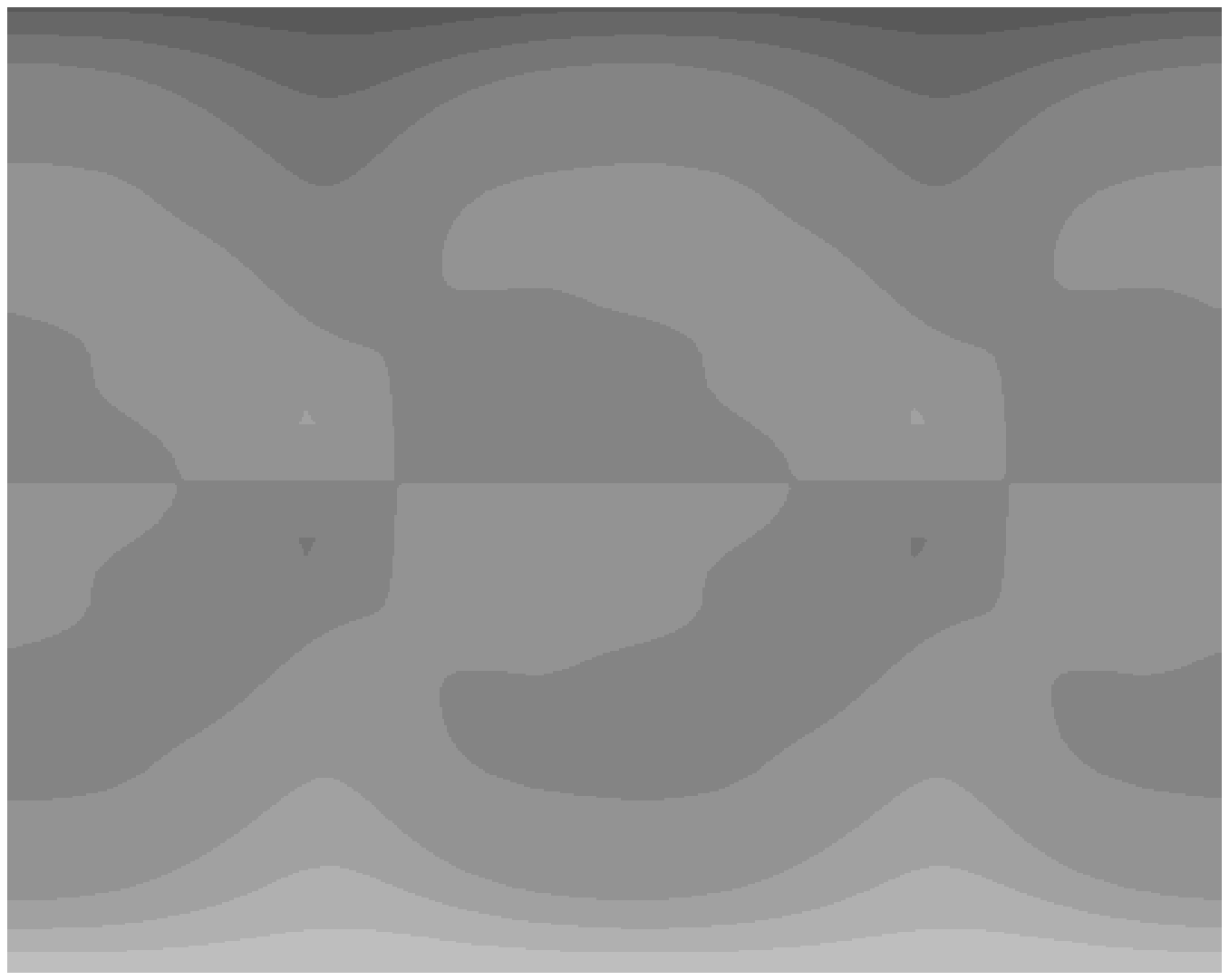, scale=0.35} 
      $0.35\,T$ 
   \end{tabular}
   \caption{ \label{fig:D20surfBr}
      $B_r$ at the surface during the reversal in Fig.\ \ref{fig:D20mnrg}.
      Patches of reversed flux near the equator 
      migrate polewards, replacing the flux at high latitudes
      with reversed field.
   }
\end{figure}

\begin{figure}
   \begin{tabular}{cc}
      $0.20\,T$
      \epsfig{figure=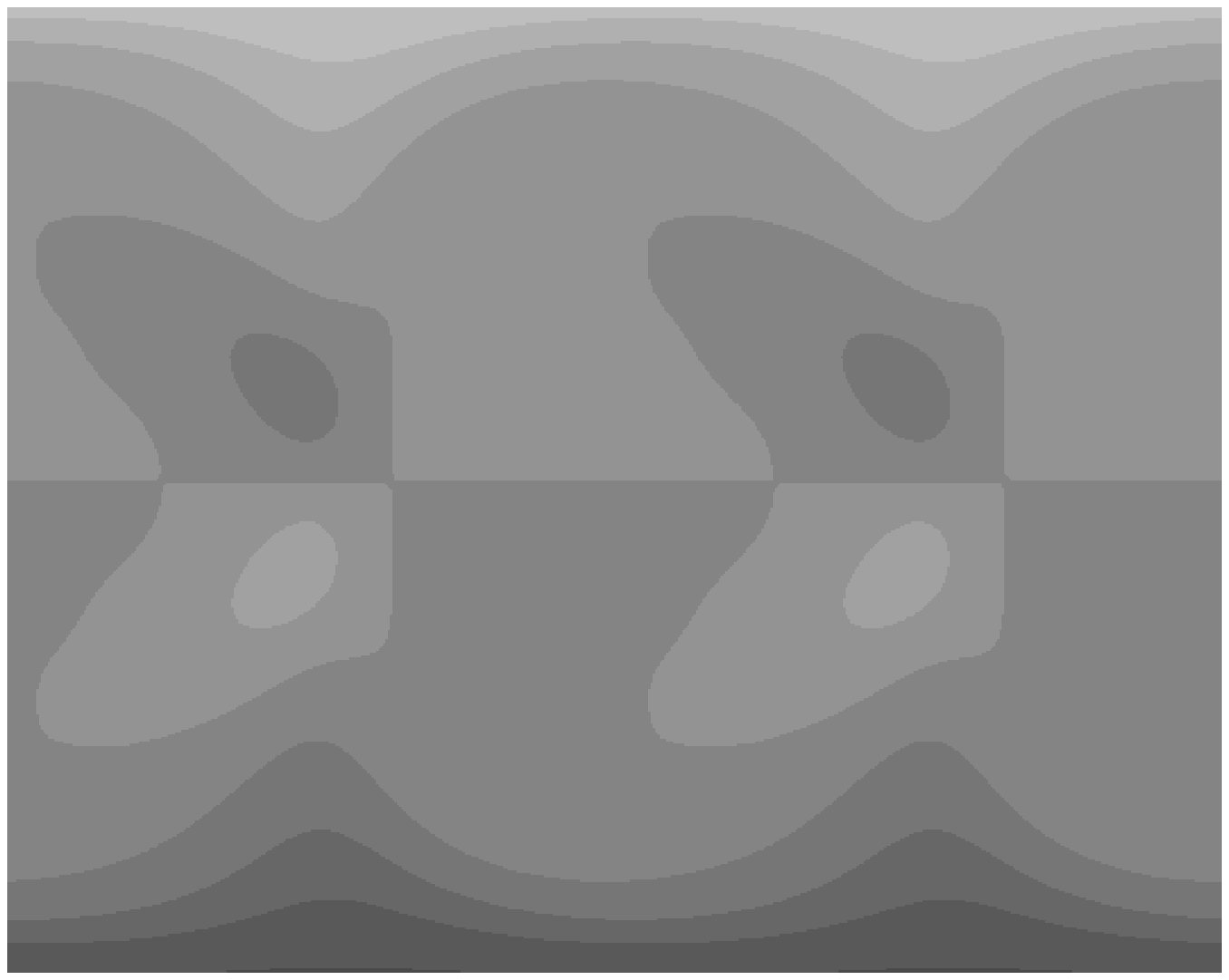, scale=0.35} &
      \epsfig{figure=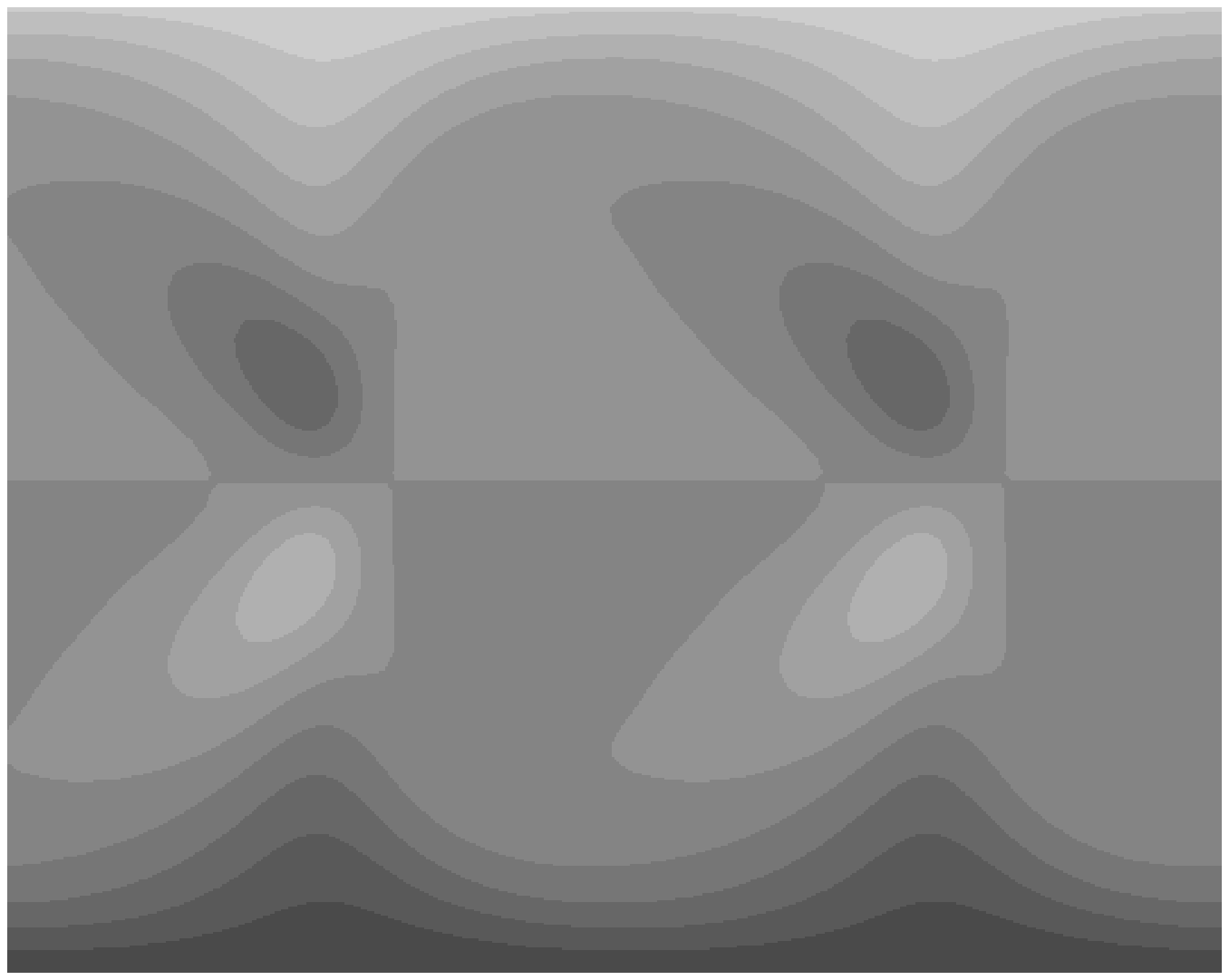, scale=0.35}
      $0.35\,T$ \\
      $0.50\,T$
      \epsfig{figure=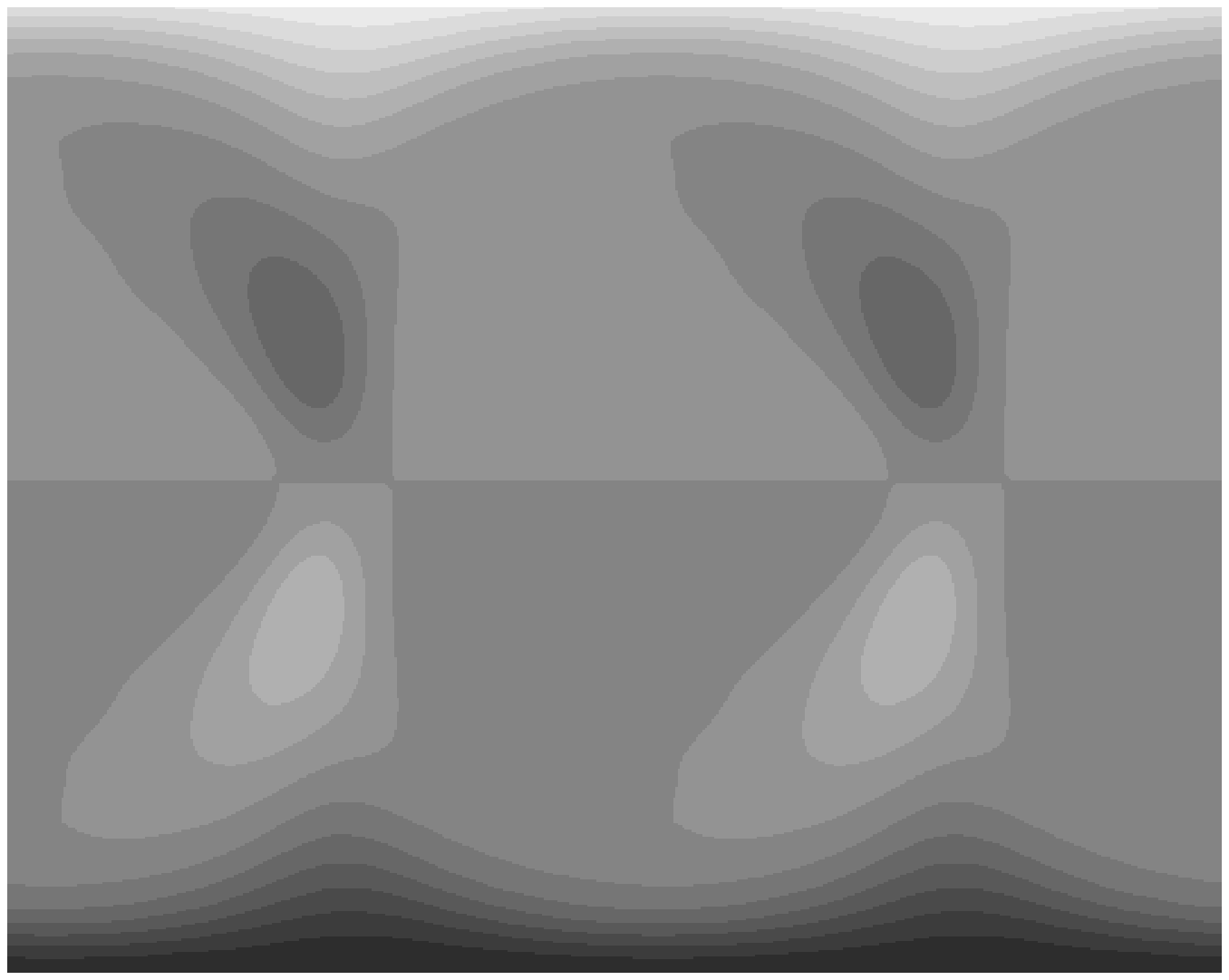, scale=0.35} &
      \epsfig{figure=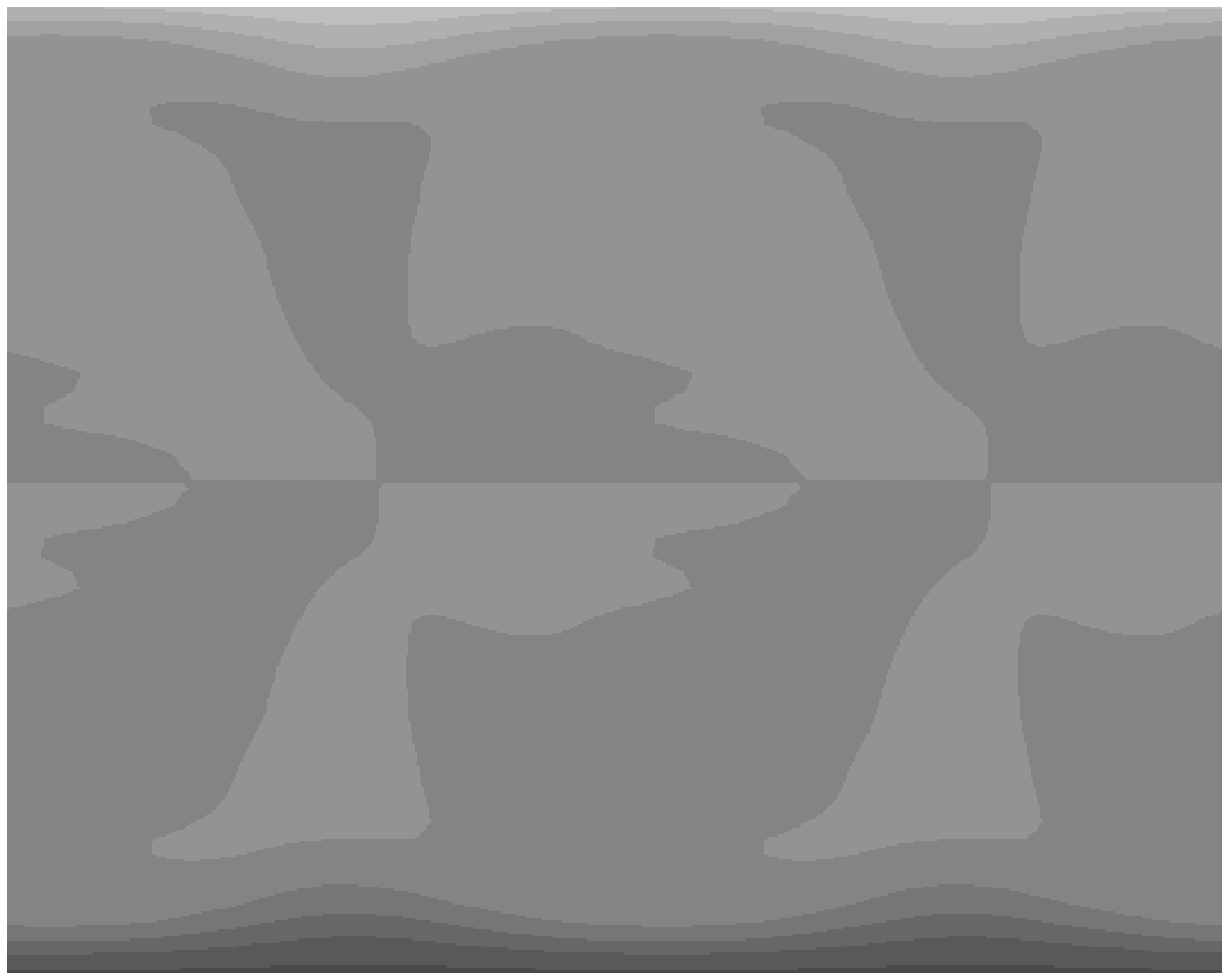, scale=0.35} 
      $0.65\,T$
   \end{tabular}
   \caption{ \label{fig:D20surfBrf}
      $B_r$ at the surface during the failed reversal 
      in Fig.\ \ref{fig:D20mnrg}.
      Reversed flux patches have insufficient time to migrate polewards
      before dissipating.
   }
\end{figure}

The reversal sequence
for $B_r$ at the surface is shown in Fig.\ \ref{fig:D20surfBr}.
Patches of reversed flux appear at low latitudes, strengthen and
migrate polewards replacing the flux at high latitudes.  
The reversal looks very similar to that reported by \citet{gubbins94}
who found the pole paths during the reversal correlate
well with the longitudes of these flux patches, located 180$^\circ$
apart.
If the frequency of the oscillation is too high the field is simply
disrupted by the short period in the oscillatory region, as seen in the 
energy in Fig.\ \ref{fig:D20mnrg}.  
This may lead to the type behaviour seen in geomagnetic excursions.
Figure \ref{fig:D20surfBrf} shows that reversed patches emerge but
have insufficient time to migrate polewards before dissipating.
They still still weaken the dipole, however.

\section{Conclusions}
We have devised a new matrix-free Krylov subspace approach to solving
the time-dependent stability problem that is most effective in exploring 
kinematic dynamo action of periodic flows. 
It is computationally efficient, uses far less
storage than conventional methods, and requires rather little new coding
once time-step and eigenvalue routines are available.

Time variation of the flow can sometimes, but not always, enhance dynamo
action. At low frequency the growth rate of the time dependent flow approaches
the average growth rate for the steady flows along the orbit. At moderate 
frequency the time dependent flow can smooth out any concentrations of
magnetic flux generated by the component steady flows. This can produce 
enhanced dynamo action (higher growth rate than the average) if the flux
concentrations are isolated and of one sign. Dynamo action is possible 
at certain frequencies even when the average growth rate for steady flows
around the cycle is negative and the cycle contains mainly steady
flows that do not generate magnetic field. The growth rate appears to 
be capped by the highest growth rate of any steady flow on the cycle.
At high frequency the magnetic field does not have time to adjust to time changes 
in the flow and becomes almost stationary. 

Time variation does not always enhance dynamo action. When the generated field
has flux concentrations of different signs close together, the smoothing 
effect tends to destroy flux.
The dynamo enhancement for these large scale flows is not as dramatic as those
reported by \citet{gog99}. Flows with similar eigenfunctions tend to have
similar growth rates at the same $R_m$, limiting the effect of the time
variation.  Where the eigenfunction changes dramatically dynamo action
is usually impaired unless the period of the flow is short compared with
the diffusion time. 


Solutions have been found that reverse when $M$ is low and are
associated with the steady-flow
oscillatory solutions found in II.
This result is in common with the reversals studied by 
\citet{sarson99}, which occur irregularly due to a 
drop in meridional circulation.  
Although flows in their calculations 
are also predominantly equatorial antisymmetric,
a small but increased flow across the equator is observed 
during a reversal.  It is unclear that this results in sufficient 
advection of flux to influence their reversal mechanism.  
A large circulation exterior to the tangent cylinder is required 
in the reversal mechanism studied by \citet{wicht04}.
It is needed to transport reversed flux originating
from plumes that protrude the tangent cylinder.
However, it is difficult 
to decipher what part fluctuations would play in their model, and in 
particular to what degree this would affect the quasi-regularity of 
their reversals.
%
We have shown that for a reversal to occur 
the drop in $M$ must persist long enough for the field to reverse, 
which for this class of flows this is approximately a tenth of a 
diffusion time, or approximately the dipole decay time.
A significant drop in magnetic energy is observed
during the reversal. This arises because of the change in eigenfunction 
between the steady and oscillatory modes of the steady solutions.
%
%

{\bf Acknowledgements} \\
 This work was supported by NERC grant GR3/12825. We thank Johannes Wicht
 and an anonymous referee for useful comments, and  
 Dr SJ Gibbons for advice on using his time-step code.

\bibliographystyle{chicago}

\newpage
\linespread{1.0}

\begin{table}
   \begin{center}
   \begin{tabular}{rccc}
      ({\it a}) \\ & $N_r$=50 & 100 & 150 \\[2pt]
      \hline
    $L$=8 &    0.86410  &   0.95152  &   0.95593 \\
       12 &    0.86705  &   0.95388  &   0.95826 \\
       16 &    0.86717  &   0.95400  &   0.95838 \\[2pt]
        8 &    -36.137  &   -36.406  &   -36.420 \\
       12 &    -35.717  &   -35.959  &   -35.971 \\
       16 &    -35.704  &   -35.954  &   -35.958 \\[12pt]
      ({\it b}) \\ & $N_r$=25 & 50 & 75 \\[2pt]
      \hline 
    $L$=8 &    0.98156  &   0.94976  &  0.95479 \\
       12 &    0.99434  &   0.95211  &  0.95709 \\
       16 &    0.99446  &   0.95223  &  0.95721 \\[2pt]
        8 &    -36.239  &   -36.452  &  -36.431 \\
       12 &    -35.788  &   -36.005  &  -35.984 \\
       16 &    -35.776  &   -35.991  &  -35.971
   \end{tabular}
   \,
   \begin{tabular}{rrr}
      ({\it c}) \\
      \multicolumn{1}{c}{$k$} & 
      \multicolumn{1}{c}{$\sigma_1$} & 
      \multicolumn{1}{c}{$\sigma(t_k)$} \\[2pt]
        \hline
       10 &  -22.29762  &  81.56889 \\
       20 &   -6.12153  &  27.14633 \\
       30 &    0.79446  &   4.61364 \\
       40 &    2.24163  &  -3.58609 \\
       50 &    0.98918  &  -3.14851 \\
       60 &    0.91102  &   0.43870 \\
       70 &    0.95688  &   3.19484 \\
       80 &    0.95722  &   4.12172 \\
       90 &    0.95721  &   3.87464 \\
      100 &    0.95721  &   3.19592 \\
      120 &             &   1.91438 \\
      150 &             &   1.09479 \\
      200 &             &   0.98221 \\
      250 &             &   0.96445 \\
      300 &             &   0.95774 \\
      400 &             &   0.95735 \\
      500 &             &   0.95733
   \end{tabular}  
   \end{center}
   \caption{ \label{tbl:grtest}
      Comparison of computed growth rates for the 
      K--R flow, $(0.9834915,0.0001632689)$, at $\Rm=1000$;
      ({\it a}) leading two eigenvalues computed using the matrix
      and its LU factorisation;
      ({\it b}) eigenvalues calculated by the matrix-free method;
      ({\it c}) comparison with simple timestepping, 
      $T=0.001$, $t_k=kT$ and $N_r=75$, $L=16$.
   }
\end{table}

%

%
%

\end{document}